\documentclass[onefignum,onetabnum]{siamart190516}

\usepackage{lipsum}
\usepackage{amsfonts}
\usepackage{graphicx}
\usepackage{epstopdf}
\usepackage{algorithmic}
\ifpdf
  \DeclareGraphicsExtensions{.eps,.pdf,.png,.jpg}
\else
  \DeclareGraphicsExtensions{.eps}
\fi

\newsiamremark{remark}{Remark}
\newsiamremark{hypothesis}{Hypothesis}
\crefname{hypothesis}{Hypothesis}{Hypotheses}
\newsiamthm{claim}{Claim}

\headers{Numerical study on viscous fingering using electric fields}{M. Zhao, P. Anjos, J. Lowengrub, W. Ying, and S. Li}

\title{Numerical study on viscous fingering using electric fields in a Hele-Shaw cell\thanks{Submitted to the editors \today
\funding{This work is funded by National Science Foundation, Division of Mathematical Sciences (NSF-DMS) grants DMS-1714973, 1719960, 1763272 (third author) and DMS- 1720420 (fifth author).  The third author is funded by the Simons Foundation (594598QN) for a NSF-Simons Center for Multiscale Cell Fate Research. The third author is also partially supported by the NIH grants 1U54CA217378-01A1 for a National Center in Cancer Systems Biology at UC Irvine and P30CA062203 for the Chao Family Comprehensive Cancer Center at UC Irvine. The fourth author is supported by the National Science Foundation of China grants DMS-11771290.
}}}

\author{Meng Zhao\thanks{Center for Mathematical Sciences, Huazhong University of Science and Technology, China
  (\email{mzhao9@hust.edu.cn}).}
\and Pedro Anjos\thanks{Department of Applied Mathematics, Illinois Institute of Technology, Chicago, IL, USA
  (\email{pamorimanjos@iit.edu}, \email{sli@math.iit.edu}).}
\and John Lowengrub\thanks{Department of Mathematics, University of California Irvine, Irvine, CA, USA
(\email{jslowengrub@gmail.com}).}
\and Wenjun Ying\thanks{School of Mathematical Sciences, MOE-LSC and Institute of Natural Sciences, Shanghai Jiao Tong University, Shanghai 200240, China
(\email{wying@sjtu.edu.cn}).}
\and Shuwang Li\footnotemark[3]}

\usepackage{amsopn}

\ifpdf
\hypersetup{
  pdftitle={Numerical study on viscous fingering using electric fields in a Hele-Shaw cell},
  pdfauthor={M. Zhao, P. Anjos, J. Lowengrub, W. Ying and S. Li}
}
\fi

\begin{document}

\maketitle

% REQUIRED
\begin{abstract}
We investigate the nonlinear dynamics of a moving interface in a Hele-Shaw cell subject to an in-plane applied electric field. We develop a spectrally accurate boundary integral method where a coupled integral equation system is formulated. Although the stiffness due to the high order spatial derivatives can be removed, the long-time simulation is still expensive since the evolving velocity of the interface drops dramatically as the interface expands. We remove this physically imposed stiffness by employing a rescaling scheme, which accelerates the slow dynamics and reduces the computational cost. Our nonlinear results reveal that positive currents restrain finger ramification and promote overall stabilization of patterns. On the other hand, negative currents make the interface more unstable and lead to the formation of thin tail structures connecting the fingers and a small inner region. When no flux is injected, and a negative current is utilized, the interface tends to approach the origin and break up into several drops. We investigate the temporal evolution of the smallest distance between the interface and the origin and find that it obeys an algebraic law $\displaystyle (t_*-t)^b$, where $t_*$ is the estimated pinch-off time.
\end{abstract}

% REQUIRED
\begin{keywords}
Hele-Shaw problem, fingering instabilities, electro-osmotic flow, boundary integral method, rescaling idea
\end{keywords}

% REQUIRED
\begin{AMS}
  45B05, 35R37, 76D27,76S05,76W05
 \end{AMS}

\section{Introduction}

Interfacial instabilities are ubiquitous in nature and engineering, such as dendritic growth in solidification \cite{Langer80,Jacob90,Sethian92,Provatas98,ShuwangDendrite}, fractal growth of diffusion-limited aggregation \cite{Witten81,Sander86}, electrodeposition of metals \cite{Brady84,Bai16}, variety patterns of tissue \cite{cristini2009,Puliafito12,Tarle15,Alert19,MJ2020}, viscous fingering in Hele-Shaw cells \cite{Saff58,Chuoke,Park,Tanveer03,Xie03,ShuwangJCP,Dallaston12,Morrow19}, and so on. In many cases, these instabilities are not expected. For instance, viscous fingering results in trapping of oil in the reservoir, thus leading to poor oil recovery \cite{Gorell83,Stokes86}; dendritic growth leads to safety problems in rechargeable batteries \cite{Xu14}; vascular tumor induces difficulty in clinic treatment \cite{Chaplain95,Macklin}. The interfacial instability has been an attractive topic for decades.

Viscous fingering in the Hele-Shaw cell, where a small gap separates two parallel plates, can be used as a prototype for investigation of interfacial instabilities \cite{HS}. When a less viscous fluid is injected into the cell and displaces a more viscous fluid, the interface separating the two fluids experiences the well-known Saffman-Taylor instabilities and as a consequence, viscous fingering patterns are formed \cite{Saff58,Langer89,Cummins}. As the interface expands, these fingers split at their tips, generating new fingers. Finally, the interface performs dense branching morphologies \cite{Chuoke,McLean,Park,Jacob86,Praud05,ShuwangJCP}. 

Recently, one variant of the classical Hele-Shaw set-up that catches researchers' attention is the Hele-Shaw problem coupled with an applied electric field \cite{Mirzadeh17,Gao19}.  In this set-up, an external electric field is utilized to promote electro-osmotic flow, which is added to the pressure-driven flow. In \cite{Mirzadeh17}, Mirzadeh and Bazant carried out a linear stability analysis of the problem in a Hele-Shaw channel, and found that the interfacial instabilities could be fully suppressed by employing sufficiently large currents.
In \cite{Gao19}, Gao {\it et al.} analyzed the same problem but in a radial Hele-Shaw displacement, where the electro-osmotic flow can either oppose or assist the pressure-driven flow. By carrying out various experiments, they also found that the electric current is able to actively control the emergence of interfacial instabilities. Although the authors provided a very clear physical explanation for the observed phenomenon, their results are limited (by their power source) to small values of electric current, which do not permit a systematic analysis of the effects of the electric field on radial viscous fingering. In addition, the experiments performed in \cite{Gao19}  are focused on the onset of the pattern formation, where the size of the fingers are still small and nonlinear effects are not significant. Therefore, the nonlinear mechanisms that dictate growth of instabilities and pattern formation are still not fully understood, especially for long times of the dynamics. Motivated by these facts, our goal in the current work is to develop an efficient and accurate numerical scheme to investigate the long-time evolution of the interface during injection-driven, electro-osmotic, radial Hele-Shaw flow.

To better understand the interfacial dynamics, we develop a boundary integral formulation to simulate the system efficiently and accurately. Herein the hydraulic and electric field are coupled leading to an integral equation system, where two double layer potentials are involved.  Unlike multiple interface problems\cite{Woods95,Daripa15,Woods15,Pedro20,Meng20,Daripa21}, both potentials are evaluated on the same boundary with different dipole densities. The densities are only needed on the interface. The boundary integral formulation reduces the two dimensional problem into one dimensional. The coupled equations can be solved via iterative methods such as GMRES \cite{GMRES}.  Once the densities are determined, the normal velocity of the interface is computed utilizing the Dirichlet-Neumann mapping \cite{LapMCD}.  The interface is updated using the second order Adams-Bashforth method where the stiffness is removed by small scale decomposition \cite{Hou94}.  Since the normal velocity of the interface decays dramatically as the interface expands, the computational cost increases and makes the long time simulations prohibited in practice.
To solve this issue, we introduce the rescaling idea where the original time and space $(\mathbf{x},t)$ is mapped into a new frame $(\bar{\mathbf{x}},\bar{t})$ \cite{ShuwangJCP,Zhao17,Meng18}. In the rescaled frame, the interface can develop at any prescribed speed while the original physics remains unchanged. Here we accelerate the interface to evolve exponentially.  A space scaling function $R(\bar t)$ is used to map the interface back to its initial size. And back to its initial size, while a time scaling function $\rho(\bar t)$ is chosen to speed up the slow dynamics. This scheme allows us to access the long time dynamics of the interface.

Our numerical simulations reveal that the method is efficient and accurate, in addition to show good agreement with linear predictions at early times. The nonlinear results demonstrate that positive currents restrain finger ramification and promote overall stabilization of patterns. Conversely, negative currents make the interface more unstable and lead to the formation of thin tail structures connecting the fingers and a small inner region. The tail region is formed by the interaction of the flux, electric current, and surface tension. When no flux is injected and a negative current is applied, we find that the interface tends to approach the origin. For lower modes, the interface does not develop thin tail and propagates rapidly to the origin. For higher modes, the interface exhibits thin tails and propagates to the origin at oscillatory speeds. The interface suggests to break up and forms drops. 
Investigating the smallest distance between the interface and the origin, we discover that the distance obeys an algebraic law $\displaystyle (t_*-t)^b$, where $t_*$ is the estimated pinch-off time and $b$ is a positive constant. Moreover, we found that $t_*$ depends on the current, surface tension, and perturbation mode.

%When the special current $I_c$ in \cref{GE:ssI} and the flux $J_d$ in \cref{GE:ssJ} is used, our simulations reveal the existence of self-similar shape.  This mode selection mechanism is the combination of the nonlinear interactions and competitions between different modes. Note the nonlinear results demonstrate that the parameters $\mathcal C$ in the current and $\mathcal D$ in the flux together play important role in this mode selection.

% The outline is not required, but we show an example here.
The paper is organized as follows. First, we present the governing equations and the linear analysis in \cref{sec:GE}; next, we investigate a rescaled boundary integral method in \cref{sec:NF}; then we discuss the numerical results in \cref{sec:Res}; and finally, we give conclusions in \cref{sec:Con}.

\section{Governing equations and linear analysis}
\label{sec:GE}
\begin{figure}[tbhp]
\centering
\includegraphics[width=0.94\textwidth]{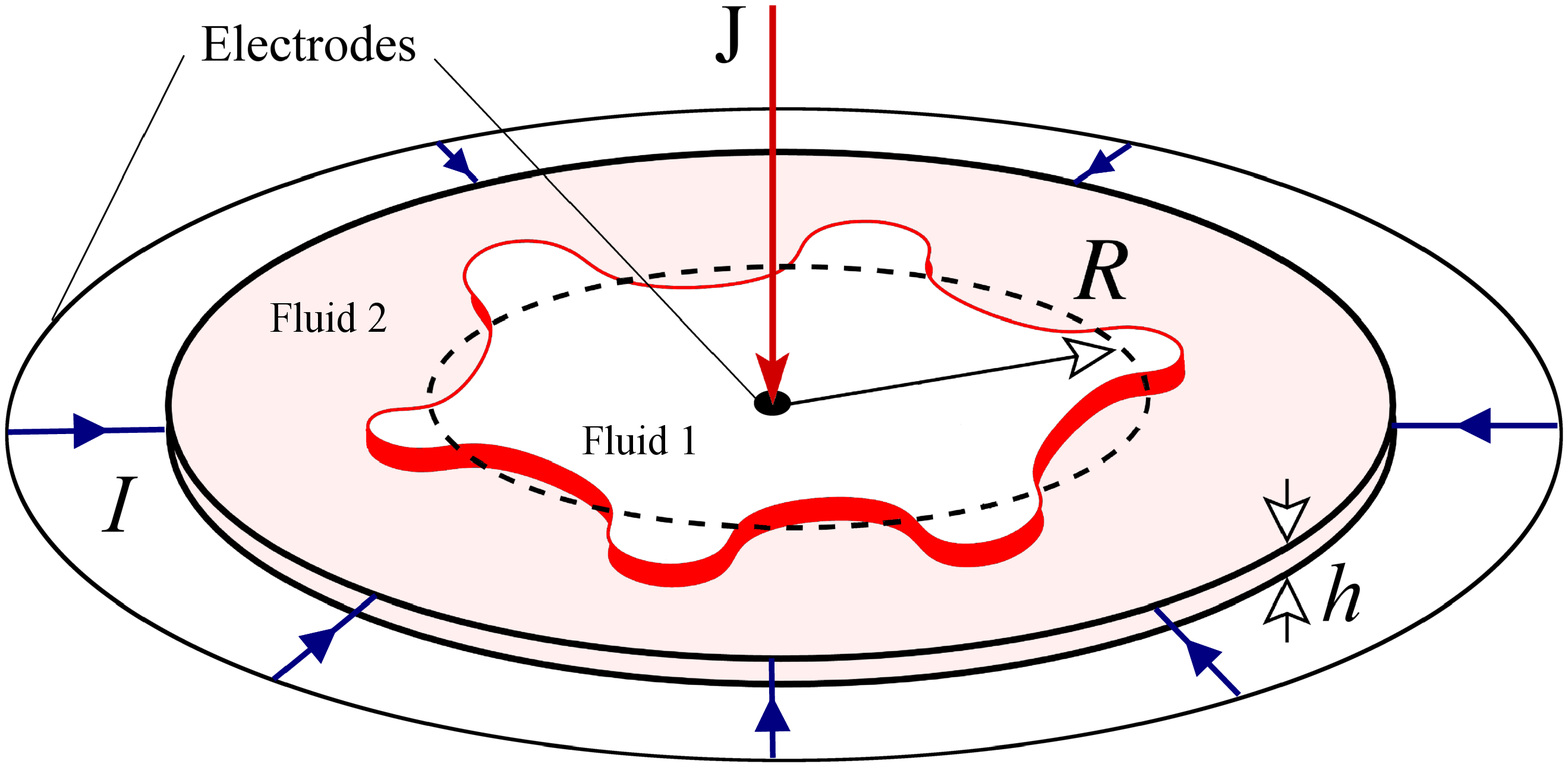}[a]
\includegraphics[scale=0.5]{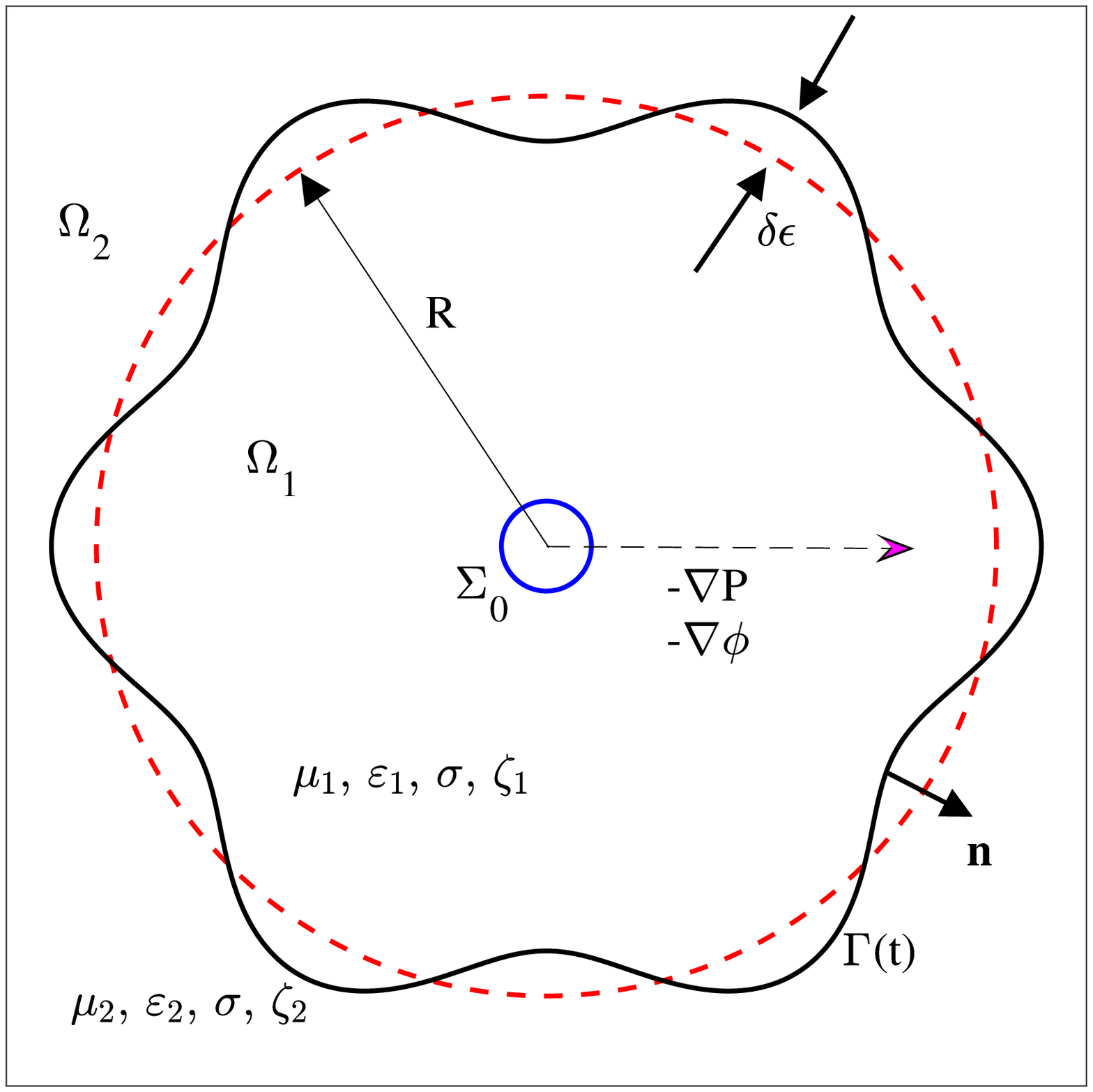}[b]
\caption{[a] shows a schematic diagram for the electrohydrodynamic Hele-Shaw flow. The inner fluid 1 is injected into an existing outer fluid 2 at a prescribed rate $J$. An electric current is produced by using the electrodes positioned at the center and far edge of the cell.  [b] represents the top view of interface as a slightly perturbed circle, which separates the two fluids.}\label{GE:fig1}
\end{figure}

We consider a radial Hele-Shaw problem \cite{Paterson81,Shelley97,Tian98,ShuwangPRL,Reis11}, where $\Gamma(t)$ is the moving interface separating the two different
fluid domains. See \cref{GE:fig1} for a schematic diagram of the Hele-Shaw cell system composed by oil (fluid 1)  and a mixture (fluid 2) of water and glycerol.  In the radial Hele-Shaw cell, the fluid 1 with viscosity $\mu_1$, permittivity $\varepsilon_1$, and zeta potential $\zeta_1$ is injected at a rate $J$ into the fluid 2 having viscosity $\mu_2$, permittivity $\varepsilon_2$, and zeta potential $\zeta_2$. An electric current $I$ is utilized through the electrodes settled at the center and the far edge of the cell.

We assume  that the hydraulic fluid flow obeys the Darcy's Law,
\begin{equation}
  \mathbf{u}_h=-k_h\nabla P,
\end{equation} 
where $\mathbf{u}_h$ is hydraulic velocity field, $P$ is pressure, and $\displaystyle k_h=\frac{h^2}{12\mu}$ is the hydraulic mobility. Here, $h$ is the cell gap width. Due to the application of an electric field, the fluids are subjected to a net electric force that drives an electro-osmotic flow,
 \begin{equation}
  \mathbf{u}_{eo}=-k_{eo}\nabla \phi,
\end{equation} 
where $\mathbf{u}_{eo}$ is the electro-osmotic velocity field, $\phi$ is the electric potential, and $\displaystyle k_{eo}=-\frac{\varepsilon\zeta}{\mu}$ is the electro-osmotic mobility. Therefore, the total velocity is the sum of hydraulic and electro-osmotic parts,
\begin{equation}
\mathbf{u}=\mathbf{u}_h+\mathbf{u}_{eo}.\label{GE:vel}
\end{equation}

As the electric field acts to generate the electro-osmotic flow, pressure gradients also influence the motion of the ions in the electric double layer (EDL) \cite{Kirby04}. This EDL is a thin region formed by an accumulation of ions in the liquids attracted by the charged surface of the glass plates of the Hele-Shaw. Therefore, the advection of ions promoted by pressure gradients results in the emergence of streaming current $\mathbf{i}_{sc}$, which is added to the Ohmic current $\mathbf{i}_{e}$ to compose the total current,
\begin{equation}
\mathbf{i}=\mathbf{i}_{sc}+\mathbf{i}_{e}=-k_{eo}\nabla P-k_e\nabla \phi,\label{GE:cur}
\end{equation}
where $k_e=\sigma$ is the  electrical conductivity. Note that the material coefficients, i.e., $\mu$, $\varepsilon$, $\zeta$ and $\sigma$, are taken in each fluid. Furthermore, subscripts 1 and 2 represent the inner and outer fluids, respectively.

We consider the fluid is incompressible and charge-free,
\begin{equation}
\nabla\cdot \mathbf{u}=0,\quad\nabla\cdot \mathbf{i}=0\label{GE:free}.
\end{equation}
Combining Eq. \cref{GE:vel,GE:cur,GE:free}, we note that $P$ and $\phi$ satisfy Laplace's equations
\begin{equation}
\nabla^2 P=0,\quad \nabla^2 \phi=0,
\end{equation}
respectively, and therefore, $P$ and $\phi$ are harmonic functions.
At the interface $\Gamma(t)$, the pressure has a jump due to the surface tension $\tau$, which is given by the well-known Young-Laplace condition. On the other hand, the electric potential $\phi$ is continuous across the interface. Therefore,
\begin{equation}
{[}P{]}=\tau\kappa, \quad [\phi]=0,
\end{equation}
where $\kappa$ is the curvature of the planar interfacial curve and the notation $[\varphi]=\varphi_1-\varphi_2$ represents the jump of a variable $\varphi$ across the interface.
In addition to these boundary conditions, we also have that the normal component of the velocity $\mathbf u$ and current $\mathbf i$ are continuous across the interface,
\begin{equation}
{[}\mathbf{u}\cdot\mathbf{n}{]}=0,\quad [\mathbf{i}\cdot\mathbf{n}]=0,
\end{equation}
where $\mathbf n$ denotes the unit vector normal to the interface $\Gamma(t)$ and pointing outward.
To complete the mathematical description of the system, we assume that the flux $J$ and total current $I$ is injected at the origin. Thus,
\begin{equation}
\int_{\Gamma_0}\mathbf{u}\cdot\mathbf{n}ds=2\pi J,\quad \int_{\Gamma_0}\mathbf{i}\cdot\mathbf{n}ds=2\pi I,
\end{equation}
where $s$ is the arclength and $\Gamma_0$ is a small circle centered at the origin.

We nondimensionalize the system using a characteristic length $L_0$, flux $J_0$, time $\displaystyle T_0=\frac{L_0^2}{J_0}$, and current $i_0$, where $L_0 = R(0)$ is the initial radius of the inner fluid region, i.e., the initial position of the unperturbed interface. Consequently, we obtain that the pressure  is scaled by $\displaystyle P_0=\frac{12\mu_2 L_0^2}{T_0h^2}$ and the electric potential is scaled by $\displaystyle \phi_0=\frac{P_0L_0}{T_0i_0}$. We also define a nondimensional surface tension $\displaystyle \tilde{\tau}=\frac
{\tau h^2T_0}{12\mu_2 L_0^3}$. The nondimensional system is given below (without changing the notations):
\begin{align}
\nabla^2 P=0,\quad \nabla^2 \phi=0 &\quad\text{for} \quad\mathbf{x}\in \Omega_1,\Omega_2,\\
{[}P{]}=\tau\kappa,\quad [\phi]=0& \quad\text{for}\quad \mathbf{x}\in\Gamma(t),\label{GE:dbc}\\
{[}\mathbf{u}\cdot\mathbf{n}{]}=0,\quad [\mathbf{i}\cdot\mathbf{n}]=0& \quad\text{for}\quad \mathbf{x}\in\Gamma(t),\label{GE:cbc} \\
\int_{\Gamma_0}\mathbf{u}\cdot\mathbf{n}ds=2\pi J,\quad \int_{\Gamma_0}\mathbf{i}\cdot\mathbf{n}ds=2\pi I.&
\end{align}

{\bf Linear stability.} In this part, we perform a linear analysis for a slightly perturbed circular interface, whose position is described by $\displaystyle r(\theta,t)=R(t)+\epsilon\delta(t)\cos(n\theta)$, where $R(t)$ is the time-dependent unperturbed radius, $n\geq 2$ is an integer perturbation mode, $\theta\in [0,2\pi]$ is the polar angle, and $\delta(t)$ is the perturbation amplitude with $\epsilon \ll 1$. We assume the generalized potential $\displaystyle \mathbf{\Phi}_i=(P_i,\phi_i)^T=\mathbf{\Phi}_i^0+\epsilon\mathbf{\Phi}_i^1+\mathcal{O}(\epsilon^2)$, where $i=1$ and $2$ represents the inner and outer fluids, respectively.
At $\mathcal{O}(1)$, we are able to find 
\begin{eqnarray}
\mathbf{\Phi}_1^0&=&-{\bf K}_1^{-1}(J, I)^T\log \frac{r}{R}+(\frac{\tau}{R},0)^T,\\
\mathbf{\Phi}_2^0&=&-{\bf K}_2^{-1}(J, I)^T\log \frac{r}{R},
\end{eqnarray}
where $\displaystyle {\bf K}_1=\left(\begin{matrix}
k_{h_1}&k_{eo_{1}}\\
k_{eo_1}&k_{e_1}
\end{matrix}\right)$ and $\displaystyle {\bf K}_2=\left(\begin{matrix}
k_{h_2}&k_{eo_{2}}\\
k_{eo_2}&k_{e_2}
\end{matrix}\right)$.

At $\mathcal{O}(\epsilon)$, we assume 
\begin{eqnarray}
\mathbf{\Phi}_1^1&=&(\beta_1,\beta_2)^T(\frac{r}{R})^n\cos(n\theta),\\
\mathbf{\Phi}_2^1&=&(\bar\beta_1,\bar\beta_2)^T(\frac{R}{r})^n\cos(n\theta),
\end{eqnarray}
and from boundary condition \cref{GE:cbc}, it reads  $\displaystyle{\bf K}_1(\beta_1, \beta_2)^T=-{\bf K}_2(\bar\beta_1,\bar\beta_2)^T$.\\
By applying \cref{GE:dbc}, we find 
\begin{equation}
(\beta_1,\beta_2)^T=\frac{\tau(n^2-1)\delta}{R^2}({\bf K}_1+{\bf K}_2)^{-1}{\bf K}_2(1,0)^T-\frac{\delta}{R}({\bf K}_1+{\bf K}_2)^{-1}{\bf K}_2({\bf K}_2^{-1}-{\bf K}_1^{-1})(J, I)^T.
\end{equation}

By using the normal velocity of the interface in the linear regime, $\displaystyle \dot{R}+\epsilon\dot{\delta}\cos(n\theta)=-k_{e_1}\frac{\partial P_1}{\partial r}-k_{eo_1}\frac{\partial \phi_1}{\partial r}$, we obtain
\begin{align}
\label{GE:Rdot}\dot{R}&=\frac{J}{R},\\
\label{GE:ddot}\dot{\delta}&=-\frac{J\delta}{R^2}-(1,0){\bf K}_1(\beta_1,\beta_2)^T\frac{n}{R},\\
\frac{\dot{\delta}}{\delta}&=\frac{nI}{R^2}(1,0){\bf K}_1({\bf K}_1+{\bf K}_2)^{-1}{\bf K}_2({\bf K}_2^{-1}-{\bf K}_1^{-1})(0, 1)^T\nonumber\\
&+\frac{nJ}{R^2}(1,0)({\bf K}_1({\bf K}_1+{\bf K}_2)^{-1}{\bf K}_2({\bf K}_2^{-1}-{\bf K}_1^{-1})-\frac{1}{n}{\bf I})(1, 0)^T\nonumber\\
&-\frac{\tau n(n^2-1)}{R^3}(1,0){\bf K}_1({\bf K}_1+{\bf K}_2)^{-1}{\bf K}_2(1,0)^T,\nonumber\\
\label{GE:drdot}&=\frac{(k_{eo_2}k_{h_1}-k_{eo_1}k_{h_2})}{(k_{eo_1}+k_{eo_2})^2-(k_{e_1}+k_{e_2})(k_{h_1}+k_{h_2})}\frac{2nI}{R^2}\\
&+(n\frac{k_{eo_1}^2-k_{eo_2}^2-(k_{e_1}+k_{e_2})(k_{h_1}-k_{h_2})}{(k_{eo_1}+k_{eo_2})^2-(k_{e_1}+k_{e_2})(k_{h_1}+k_{h_2})}-1)\frac{J}{R^2}\nonumber\\
&-\frac{\tau n(n^2-1)}{R^3}\frac{(k_{eo_2}^2k_{h_1}+(k_{eo_1}^2-(k_{e_1}+k_{e_2})k_{h_1})k_{h_2})}{(k_{eo_1}+k_{eo_2})^2-(k_{e_1}+k_{e_2})(k_{h_1}+k_{h_2})}.\nonumber
\end{align}

We introduce the quantity $\displaystyle \frac \delta R (t)=\frac{\delta(t)}{R(t)}$, which is known as the shape factor and characterizes the size of the perturbation relative to the underlying circle. By utilizing Eqs. \cref{GE:Rdot,GE:ddot,GE:drdot}, one verifies that the
shape factor evolves according to
\begin{multline}
(\frac{\delta}{R})^{-1}\frac{d}{dt}(\frac{\delta}{R})=\frac{(k_{eo_2}k_{h_1}-k_{eo_1}k_{h_2})}{(k_{eo_1}+k_{eo_2})^2-(k_{e_1}+k_{e_2})(k_{h_1}+k_{h_2})}\frac{2nI}{R^2}\\
+(n\frac{k_{eo_1}^2-k_{eo_2}^2-(k_{e_1}+k_{e_2})(k_{h_1}-k_{h_2})}{(k_{eo_1}+k_{eo_2})^2-(k_{e_1}+k_{e_2})(k_{h_1}+k_{h_2})}-2)\frac{J}{R^2}\\
-\frac{\tau n(n^2-1)}{R^3}\frac{(k_{eo_2}^2k_{h_1}+(k_{eo_1}^2-(k_{e_1}+k_{e_2})k_{h_1})k_{h_2})}{(k_{eo_1}+k_{eo_2})^2-(k_{e_1}+k_{e_2})(k_{h_1}+k_{h_2})}.\label{GE:sf}
\end{multline}

When there is no electric field, that is, $I=0$ and $k_{eo_1}=k_{eo_2}=0$, equation \cref{GE:sf} reduces to the classical hydraulic form \cite{ShuwangJCP}.  Note that surface tension $\tau$ suppresses instabilities of the interface. The flux $J$ promotes the interfacial instability when a less viscous fluid is injected into a more viscous fluid. But the current $I$ is able to have both effects depending on the direction, which is related to its sign. Thus, the stability of the interface is the balance of these three terms. In other words, we are able to control the interfacial stability by only manipulating the current $I$ \cite{Mirzadeh17,Gao19}.

Taking the derivatives of  Eq. \cref{GE:sf} with respect to $n$ and setting it to be  zero, we are able to find the fastest growing mode 
\begin{multline}
n_{max}=\sqrt{\frac 1 3\left(\frac{(2I(k_{eo_2}k_{h_1}-k_{eo_1}k_{h_2})+J[k_{eo_1}^2-k_{eo_2}^2-(k_{e_1}+k_{e_2})(k_{h_1}-k_{h_2})])}{\tau[k_{eo_2}^2k_{h_1}+(k_{eo_1}^2-(k_{e_1}+k_{e_2})k_{h_1})k_{h_2}]}R+1\right)}.
\end{multline}
By employing a time-dependent current
\begin{multline}
I_c=\bigg(-[k_{eo_1}^2-k_{eo_2}^2-(k_{e_1}+k_{e_2})(k_{h_1}-k_{h_2})]J\\
+\frac{\tau \mathcal{C}}{R}[k_{eo_2}^2k_{h_1}+(k_{eo_1}^2-(k_{e_1}+k_{e_2})k_{h_1})k_{h_2}]\bigg)/2/(k_{eo_2}k_{h_1}-k_{eo_1}k_{h_2}),\label{GE:ssI}
\end{multline}
where $\mathcal{C}$ is a constant, $n_{max}$ becomes independent of $R$ and thus fixed in time as the interface grows. Specifically, when $\mathcal{C}=3n_{max}^2-1$, the fastest growing mode $n_{max}$ is prescribed and sets the symmetry of the interface.  Note that once $\mathcal C$ is chosen, the fastest growing mode remains unchanged regardless of the employed flux $J$.

Using the current in  Eq. \cref{GE:ssI}, the fastest growing mode $n_{max}$ has the following growth rate,
\begin{equation}
(\frac{{\delta}}{R})^{-1}\frac{d}{d t}(\frac{{\delta}}{R})=\frac{2n_{max}^3\tau}{R^3}\frac{[k_{eo_2}^2k_{h_1}+(k_{eo_1}^2-(k_{e_1}+k_{e_2})k_{h_1})k_{h_2}]}{(k_{eo_1}+k_{eo_2})^2-(k_{e_1}+k_{e_2})(k_{h_1}+k_{h_2})}-\frac{2J}{R^2}.\label{GE:ssg}
\end{equation}
Note that when $J=0$, the area of the interior fluid does not change and $R(t)=1$. As a result, $I_c$ becomes a constant. When $J$ is a positive constant, it is easy to see that \cref{GE:ssg} is dominated by the flux term (the second term). In this case, the growth rate is negative for large interfacial sizes, indicating that the interface eventually becomes a circle. On the other hand, if we choose $J\sim R^{-1}$, these two terms in  \cref{GE:ssg} scale with $R^{-3}$. Particularly, we take
 \begin{equation}
 J_d=\frac{\tau \mathcal{D}}{R}\frac{k_{eo_2}^2k_{h_1}+(k_{eo_1}^2-(k_{e_1}+k_{e_2})k_{h_1})k_{h_2}}{k_{eo_1}^2-k_{eo_2}^2-(k_{e_1}+k_{e_2})(k_{h_1}-k_{h_2})}, \label{GE:ssJ}
 \end{equation} 
 where $\mathcal{D}$ is a constant. When $\mathcal{D}=\mathcal{C}$, the current is zero. It is again reduced to the classical hydraulic case.

\section{Numerical formulations}
\label{sec:NF}
\subsection{Boundary integral formulation}
We consider two potentials, $\displaystyle \varphi_{v_i}=k_{h_i}P_i+k_{eo_i}\phi_i$ and $\displaystyle \varphi_{c_i}=k_{eo_i}P_i+k_{e_i}\phi_i$, which are harmonic functions. Suppose that they have the form as double layer potentials, 
\begin{eqnarray}
\varphi_{v}&=&\frac{1}{2\pi}\int_{\Gamma(t)} \gamma_1(\mathbf{y})\frac{\partial\ln |\mathbf{x}-\mathbf{y}|}{\partial \mathbf{n(y)}}ds(\mathbf{y})+J\ln|\mathbf{x}|,\\
\varphi_{c}&=&\frac{1}{2\pi}\int_{\Gamma(t)} \gamma_2(\mathbf{y})\frac{\partial\ln |\mathbf{x}-\mathbf{y}|}{\partial \mathbf{n(y)}}ds(\mathbf{y})+I\ln|\mathbf{x}|,
\end{eqnarray}
where $\gamma_1$ and $\gamma_2$ are dipole densities on the interface. The continuous boundary conditions \cref{GE:cbc} are satisfied automatically. From the boundary conditions \cref{GE:dbc}, we have the following integral equations,
\begin{multline}
(\frac{k_{e_1}}{k_{h_1}k_{e_1}-k_{eo_1}^2}+\frac{k_{e_2}}{k_{h_2}k_{e_2}-k_{eo_2}^2})\gamma_1+\frac{1}{\pi}(\frac{k_{e_1}}{k_{h_1}k_{e_1}-k_{eo_1}^2}-\frac{k_{e_2}}{k_{h_2}k_{e_2}-k_{eo_2}^2})\int_{\Gamma(t)} \gamma_1(\mathbf{y})\frac{\partial\ln |\mathbf{x}-\mathbf{y}|}{\partial \mathbf{n(y)}}ds(\mathbf{y})\\
-(\frac{k_{eo_1}}{k_{h_1}k_{e_1}-k_{eo_1}^2}+\frac{k_{eo_2}}{k_{h_2}k_{e_2}-k_{eo_2}^2})\gamma_2-\frac{1}{\pi}(\frac{k_{eo_1}}{k_{h_1}k_{e_1}-k_{eo_1}^2}-\frac{k_{eo_2}}{k_{h_2}k_{e_2}-k_{eo_2}^2})\int_{\Gamma(t)} \gamma_2(\mathbf{y})\frac{\partial\ln |\mathbf{x}-\mathbf{y}|}{\partial \mathbf{n(y)}}ds(\mathbf{y})\\
=2\tau\kappa-(\frac{k_{e_1}}{k_{h_1}k_{e_1}-k_{eo_1}^2}-\frac{k_{e_2}}{k_{h_2}k_{e_2}-k_{eo_2}^2})J\ln|\mathbf{x}|^2+(\frac{k_{eo_1}}{k_{h_1}k_{e_1}-k_{eo_1}^2}-\frac{k_{eo_2}}{k_{h_2}k_{e_2}-k_{eo_2}^2})I\ln|\mathbf{x}|^2,\label{integr1}
\end{multline}
\begin{multline}
-(\frac{k_{eo_1}}{k_{h_1}k_{e_1}-k_{eo_1}^2}+\frac{k_{eo_2}}{k_{h_2}k_{e_2}-k_{eo_2}^2})\gamma_1-\frac{1}{\pi}(\frac{k_{eo_1}}{k_{h_1}k_{eo_1}-k_{eo_1}^2}-\frac{k_{eo_2}}{k_{h_2}k_{e_2}-k_{eo_2}^2})\int_{\Gamma(t)} \gamma_1(\mathbf{y})\frac{\partial\ln |\mathbf{x}-\mathbf{y}|}{\partial \mathbf{n(y)}}ds(\mathbf{y})\\
+(\frac{k_{h_1}}{k_{h_1}k_{e_1}-k_{eo_1}^2}+\frac{k_{h_2}}{k_{h_2}k_{e_2}-k_{eo_2}^2})\gamma_2+\frac{1}{\pi}(\frac{k_{h_1}}{k_{h_1}k_{e_1}-k_{eo_1}^2}-\frac{k_{h_2}}{k_{h_2}k_{e_2}-k_{eo_2}^2})\int_{\Gamma(t)} \gamma_2(\mathbf{y})\frac{\partial\ln |\mathbf{x}-\mathbf{y}|}{\partial \mathbf{n(y)}}ds(\mathbf{y})\\
=(\frac{k_{eo_1}}{k_{h_1}k_{e_1}-k_{eo_1}^2}-\frac{k_{eo_2}}{k_{h_2}k_{e_2}-k_{eo_2}^2})J\ln|\mathbf{x}|^2-(\frac{k_{h_1}}{k_{h_1}k_{e_1}-k_{eo_1}^2}-\frac{k_{h_2}}{k_{h_2}k_{e_2}-k_{eo_2}^2})I\ln|\mathbf{x}|^2.
\label{integr2}
\end{multline}
These equations are well-defined Fredholms integral equations of the second kind. We are able to solve them via iterative methods such as GMRES  \cite{GMRES}. Once the dipole densities $\gamma_1$ and $\gamma_2$ are solved, we are able to compute the normal velocity of the interface via Dirichlet-Neumann mapping \cite{LapMCD}:  
\begin{equation}
V(t)=\frac{1}{2\pi}\int_{\Gamma(t)}\gamma_{1s'}\frac{(\mathbf{x}-\mathbf{x}')^\perp\cdot\mathbf{n(x)}}{|\mathbf{x}-\mathbf{x}'|^2}ds'(\mathbf{x'})+ J\frac{\mathbf{x}\cdot\mathbf{n}}{|\mathbf{x}|^2},\label{NF:V1}
\end{equation}
where the subscript $s'$ denotes the partial derivatives with respect to arclength and ${\textbf{x}}^{\perp}=(x_2,-x_1)$. The interface is evolved through,
\begin{equation}
\frac{d{\mathbf{x}}(t,\theta)}{d t}\cdot \mathbf{n}=V( t,\theta).
\end{equation}
Note that in \cref{NF:V1}, the normal velocity decreases as the interface size $|\bf x|$ gets large. It prohibits one from computing the dynamics of an interface at long times. Thus, we introduce the following rescaling scheme to accelerate this slow dynamics.

\subsection{Rescaling idea}
Following \cite{ShuwangJCP,MengJFM, Zhao17}, we introduce 
\begin{equation}
\mathbf{x}=\bar R(\bar t)\mathbf{\bar x}(\bar t,\theta),
\label{xs}
\end{equation}
and
\begin{equation}
\bar t=\int_{0}^t \frac{1}{\rho(t^\prime)}dt^\prime,
\label{ts}
\end{equation} 
where the space scaling  $\bar R(\bar t)$ represents the size of the interface, $\bar{\textbf{x}}$ is the position vector of the scaled interface, and $\theta$ parameterizes the interface. The time scaling function $\rho(t)=\bar{\rho}(\bar{t})$ maps the original time $t$ to the new time $\bar t$ and $\rho(t)$ has to be positive and continuous. The evolution of the interface in the scaled frame can be accelerated \cite{Zhao2015,Zhao17} or decelerated \cite{Meng18,MengJFM} by choosing a different $\rho(t)$. A straightforward calculation shows  the normal velocity in the new frame
\begin{equation}
\bar{V}(\bar t)=\frac{\bar{\rho}}{\bar R}V(t(\bar t))-\frac{\mathbf{\bar x}\cdot \mathbf{n}}{\bar R}\frac{d\bar R}{d\bar t}.
\label{NF:Vrelation}
\end{equation}
In the rescaled frame, we require that the area enclosed by the interface remains constant $\bar{A}(\bar t)=\bar{A}(0)$. That is, the integration of the normal velocity along the interface in the scaled frame vanishes $\displaystyle \int_{\bar{\Gamma}(\bar t)}\bar{V}d\bar{s}=0$.
As a consequence, 
\begin{equation}
\frac{d\bar R}{d\bar t}=\frac{\pi\bar{\rho}\bar{J}}{\bar{A}(0)\bar R}.
\end{equation}
Choosing $\displaystyle \rho(\bar{t})= \bar R^2(\bar t)$, we have the interface grows exponentially in the rescaled frame,
\begin{equation}
\bar{R}(\bar{t})=\exp(\frac{\pi\bar{J}}{\bar{A}(0)}\bar t). 
\end{equation}
Taking $\displaystyle \bar \gamma_1(\mathbf{\bar x})=\gamma_1(\mathbf{x})\bar R(\bar t)$ and $\displaystyle \bar \gamma_2(\mathbf{\bar x})=\gamma_2(\mathbf{x})\bar R(\bar t)$, we next rewrite the integral equations \cref{integr1,integr2} in the rescaled frame as
\begin{multline}\label{integr12}
(\frac{k_{e_1}}{k_{h_1}k_{e_1}-k_{eo_1}^2}+\frac{k_{e_2}}{k_{h_2}k_{e_2}-k_{eo_2}^2})\bar\gamma_1+\frac{1}{\pi}(\frac{k_{e_1}}{k_{h_1}k_{e_1}-k_{eo_1}^2}-\frac{k_{e_2}}{k_{h_2}k_{e_2}-k_{eo_2}^2})\int_{\bar{\Gamma}(\bar t)} \bar\gamma_1(\bar{\mathbf{y}})\frac{\partial\ln |\bar{\mathbf{x}}-\bar{\mathbf{y}}|}{\partial \mathbf{n(\bar y)}}d\bar s(\bar{\mathbf{y}})\\
-(\frac{k_{eo_1}}{k_{h_1}k_{e_1}-k_{eo_1}^2}+\frac{k_{eo_2}}{k_{h_2}k_{e_2}-k_{eo_2}^2})\bar\gamma_2-\frac{1}{\pi}(\frac{k_{eo_1}}{k_{h_1}k_{e_1}-k_{eo_1}^2}-\frac{k_{eo_2}}{k_{h_2}k_{e_2}-k_{eo_2}^2})\int_{\bar{\Gamma}(\bar t)} \bar\gamma_2(\bar{\mathbf{y}})\frac{\partial\ln |\bar{\mathbf{x}}-\bar{\mathbf{y}}|}{\partial \mathbf{n(\bar y)}}d\bar s(\bar{\mathbf{y}})\\
=2\tau\bar\kappa-(\frac{k_{e_1}}{k_{h_1}k_{e_1}-k_{eo_1}^2}-\frac{k_{e_2}}{k_{h_2}k_{e_2}-k_{eo_2}^2})J\bar R(2\ln \bar R+\ln|\bar{\mathbf{x}}|^2)+(\frac{k_{eo_1}}{k_{h_1}k_{e_1}-k_{eo_1}^2}-\frac{k_{eo_2}}{k_{h_2}k_{e_2}-k_{eo_2}^2})I\bar R(2\ln \bar R+\ln|\bar{\mathbf{x}}|^2),
\end{multline}
\begin{multline}\label{integr22}
-(\frac{k_{eo_1}}{k_{h_1}k_{e_1}-k_{eo_1}^2}+\frac{k_{eo_2}}{k_{h_2}k_{e_2}-k_{eo_2}^2})\bar \gamma_1-\frac{1}{\pi}(\frac{k_{eo_1}}{k_{h_1}k_{eo_1}-k_{eo_1}^2}-\frac{k_{e_2}}{k_{h_2}k_{e_2}-k_{eo_2}^2})\int_{\bar{\Gamma}(\bar t)} \bar \gamma_1(\bar{\mathbf{y}})\frac{\partial\ln |\bar{\mathbf{x}}-\bar{\mathbf{y}}|}{\partial \mathbf{n(\bar y)}}d\bar s(\bar{\mathbf{y}})\\
+(\frac{k_{h_1}}{k_{h_1}k_{e_1}-k_{eo_1}^2}+\frac{k_{h_2}}{k_{h_2}k_{e_2}-k_{eo_2}^2})\bar\gamma_2+\frac{1}{\pi}(\frac{k_{h_1}}{k_{h_1}k_{e_1}-k_{eo_1}^2}-\frac{k_{h_2}}{k_{h_2}k_{e_2}-k_{eo_2}^2})\int_{\bar{\Gamma}(\bar t)} \bar\gamma_2(\bar{\mathbf{y}})\frac{\partial\ln |\bar{\mathbf{x}}-\bar{\mathbf{y}}|}{\partial \mathbf{n(\bar y)}}d\bar s(\bar{\mathbf{y}})\\
=(\frac{k_{eo_1}}{k_{h_1}k_{e_1}-k_{eo_1}^2}-\frac{k_{eo_2}}{k_{h_2}k_{e_2}-k_{eo_2}^2})J\bar R(2\ln \bar R+\ln|\bar{\mathbf{x}}|^2)-(\frac{k_{h_1}}{k_{h_1}k_{e_1}-k_{eo_1}^2}-\frac{k_{h_2}}{k_{h_2}k_{e_2}-k_{eo_2}^2})I\bar R(2\ln \bar R+\ln|\bar{\mathbf{x}}|^2).
\end{multline}
Using \cref{NF:V1}, we are able to compute the normal velocity in the rescaled frame,
     \begin{equation}
     \bar V(\mathbf{\bar x})=\frac{1}{2\pi\bar R}\int_{\bar{\Gamma}(\bar t)} \bar \gamma_{1\bar{s}}\frac{(\mathbf{\bar x}'-\mathbf{\bar x})^{\perp}\cdot\mathbf{\bar{n}}(\bar{\mathbf{x}})}{|\mathbf{\bar x}'-\mathbf{\bar x}|^2}d\bar{s}'+\bar{J}\frac{\mathbf{\bar x}\cdot \mathbf{\bar{n}}}{|\mathbf{\bar x}|^2}-\frac{\pi\bar{J}}{\bar{A}(0)}\mathbf{\bar x}\cdot \mathbf{\bar{n}},
     \label{NF:Vb}
     \end{equation}
where $\mathbf{\bar x}^{\perp}=(\bar{x}_2,-\bar x_1)$. Then we evolve the interface in the scaled frame through
   \begin{equation}
   \frac{d\bar{\textbf{x}}(\bar t,\theta)}{d\bar t}\cdot \textbf{n}=\bar V(\bar t,\theta).
   \label{EvolveRS}
   \end{equation}

 To evolve the interface numerically,  \cref{integr12,integr22} are discretized in space using spectrally accurate
 discretizations \cite{HS,ShuwangJCP}, and  the integrals in
are evaluated using the
 fast multipole method \cite{Greengard87}. The discrete system is solved efficiently using GMRES
 \cite{GMRES}. Because \cref{integr12,integr22} are well-conditioned, no
 preconditioner is needed. Once the solution to the integral equation
 is obtained, the Dirichlet-Neumann map \cite{LapMCD} is used to
 determine the normal velocity of the interface via
\cref{NF:Vb} in the scaled frame. Similar to the methods implemented in \cite{MengJFM,Amlan2013,Kara2018,MJ2019}, we then evolve the interface  in the scaled frame using a second order accurate non-stiff updating
 scheme (Adams-Bashforth type) in time and the equal arclength parameterization
 \cite{Hou94,Hou01}.  The outline of our algorithm is given as the following.
%{\color{blue} Meng, it may be a good idea to add a psudo-code here}

\begin{algorithm}
\label{alg}
\begin{algorithmic}
\STATE{Initialize the interface shape}
\WHILE{$\bar t < T$}
\STATE{Use the GMRES to solve \cref{integr12,integr22} for $\bar \gamma_1$ and $\bar \gamma_2$}
\STATE{Apply the Dirichlet-Neumann mapping to compute \cref{NF:Vb} for the normal velocity $\bar{V}$}
\STATE{Use \cref{EvolveRS} to update the interface  $\bar{\mathbf{x}}$}
\STATE{Update $\bar t$ and repeat}
\ENDWHILE
\end{algorithmic}
\end{algorithm}

\section{Numerical results}
\label{sec:Res}
\subsection{Convergence test}
In this section, we test the convergence of our scheme. We take the initial shape to be $r(\theta,0)=1+0.1\times \cos(4\theta)$. The oil is injected into the water and glycerol mixture at a constant flow rate ${J}=1$; a constant current $I=-4I_0=-636$ flows from interior to exterior;  and the surface tension is $\tau=0.0216$.  Other nondimensional parameters are $k_{h_1}=14.93$, $k_{h_2}=1$, $k_{eo_1}=0$, $k_{eo_2}=1.93\times 10^{-4}$, and $k_{e_1}=k_{e_2}=2.66$. Note that we obtain these nondimenional parameters from available experiments \cite{Gao19}.

\begin{figure}[tbhp]
\centering
\includegraphics[scale=0.28]{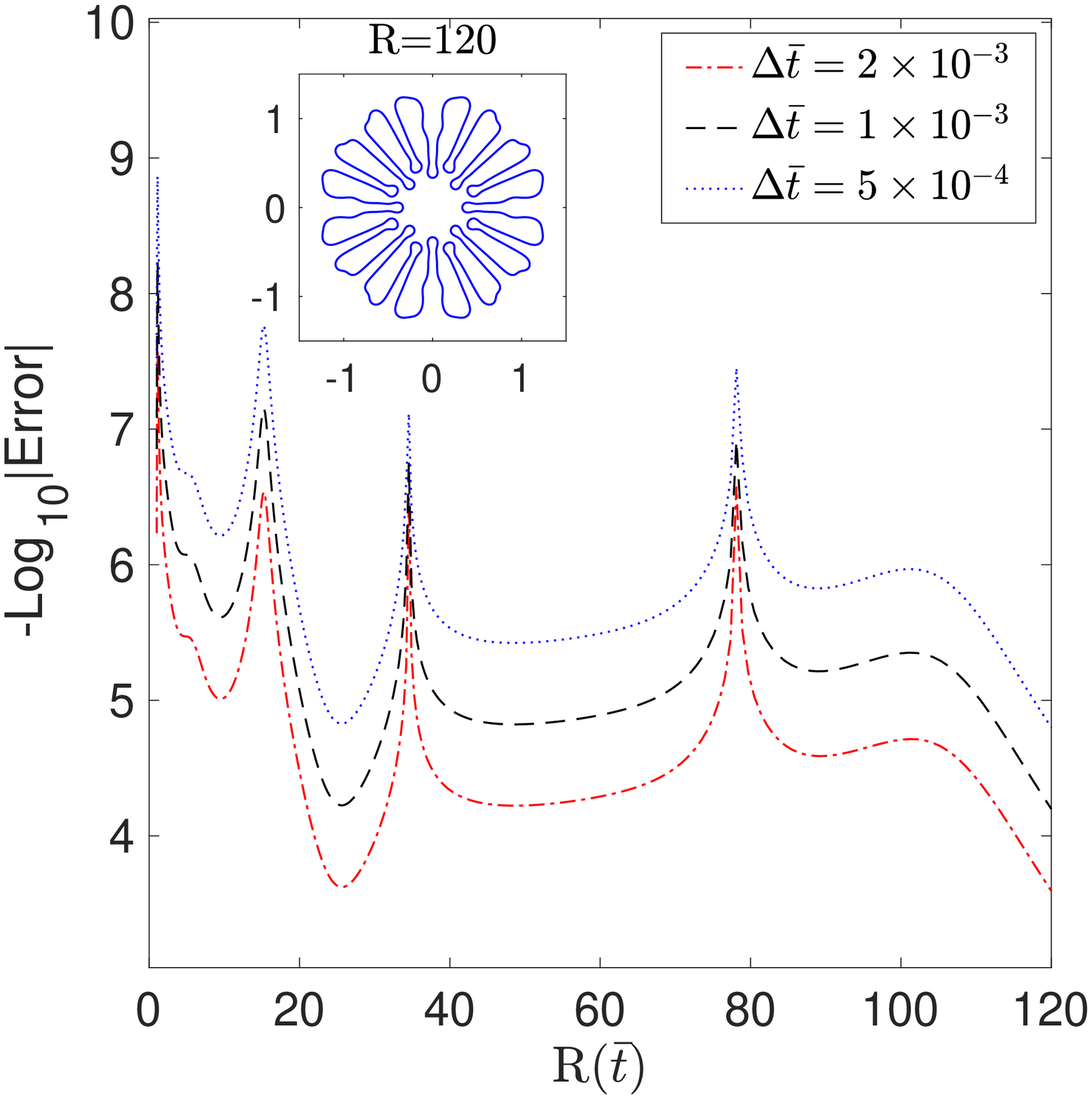}[a]
\includegraphics[scale=0.28]{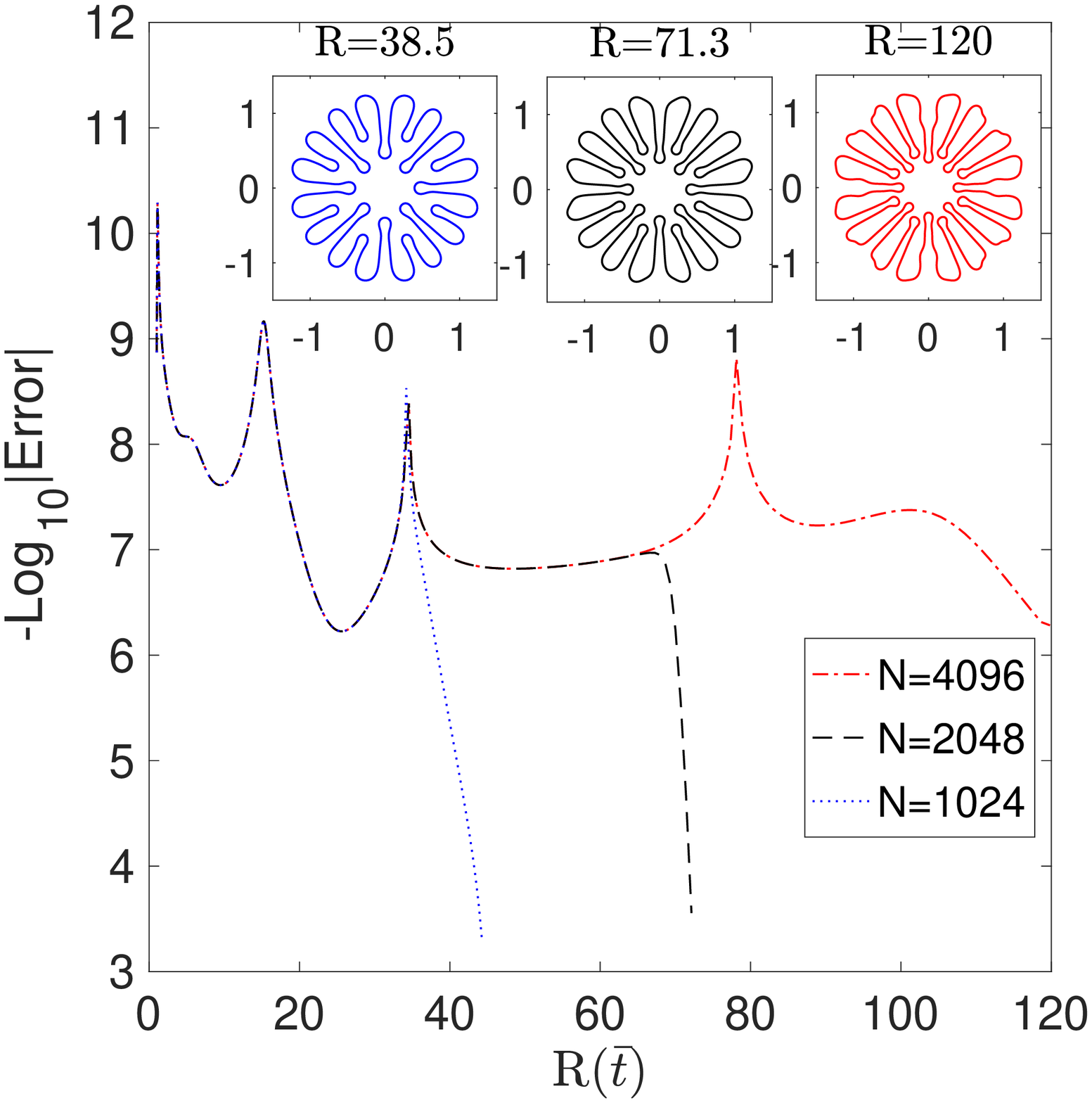}[b]
\caption{[a] shows the scheme is convergent in time. [b] shows the scheme is convergent in space.}\label{Res:fig1}
\end{figure}

First we study the temporal resolution using $N=4096$ mesh points along the interface. The time steps are set as $\Delta \bar{t}=2\times 10^{-3}$, $1\times 10^{-3}$, and $5\times 10^{-4}$. The numerical error is measured by
$Error=|\bar{A}_{\bar{t}}-\bar{A}_0|$, where $\bar{A}_{\bar{t}}$ is the area enclosed
by the interface in the scaled frame at time $\bar{t}$, and $\bar{A}_0$ is the initial
area.  \Cref{Res:fig1}[a] shows the base 10 logarithm of the temporal error plotted versus the scaling factor $R(t)=\bar{R}(\bar{t})$. The morphology of the interfaces in rescaled frame is shown as an inset. When the time step is reduced by half, the numerical error are decreased by 0.6 in distance indicating the convergent rate in time is almost 2. 

Next we study the resolution in space. We compare the shape of the interface using resolution  $N=1024$,
$2048$, $4096$, and the time step $\Delta \bar{t}=1\times 10^{-4}$. The error is again measured as $Error=|\bar{A}_{\bar{t}}-\bar{A}_0|$. The results are presented in \cref{Res:fig1}[b]. The detailed morphologies in rescaled frame are shown as insets.
 When the error is greater than $8\times 10^{-7}$, $\bar{R}(\bar{t})=38.5$ is the radius of the interface for $N=1024$; $\bar{R}(\bar{t})=71.3$ is the radius for $N=2048$; $\bar{R}(\bar{t})=120$ is the radius for $N=4096$. The morphologies at the same radius are identical. To run longer, more mesh points are needed to resolve the complicated interface.

\begin{figure}[tbhp]
\centering
\includegraphics[scale=0.28]{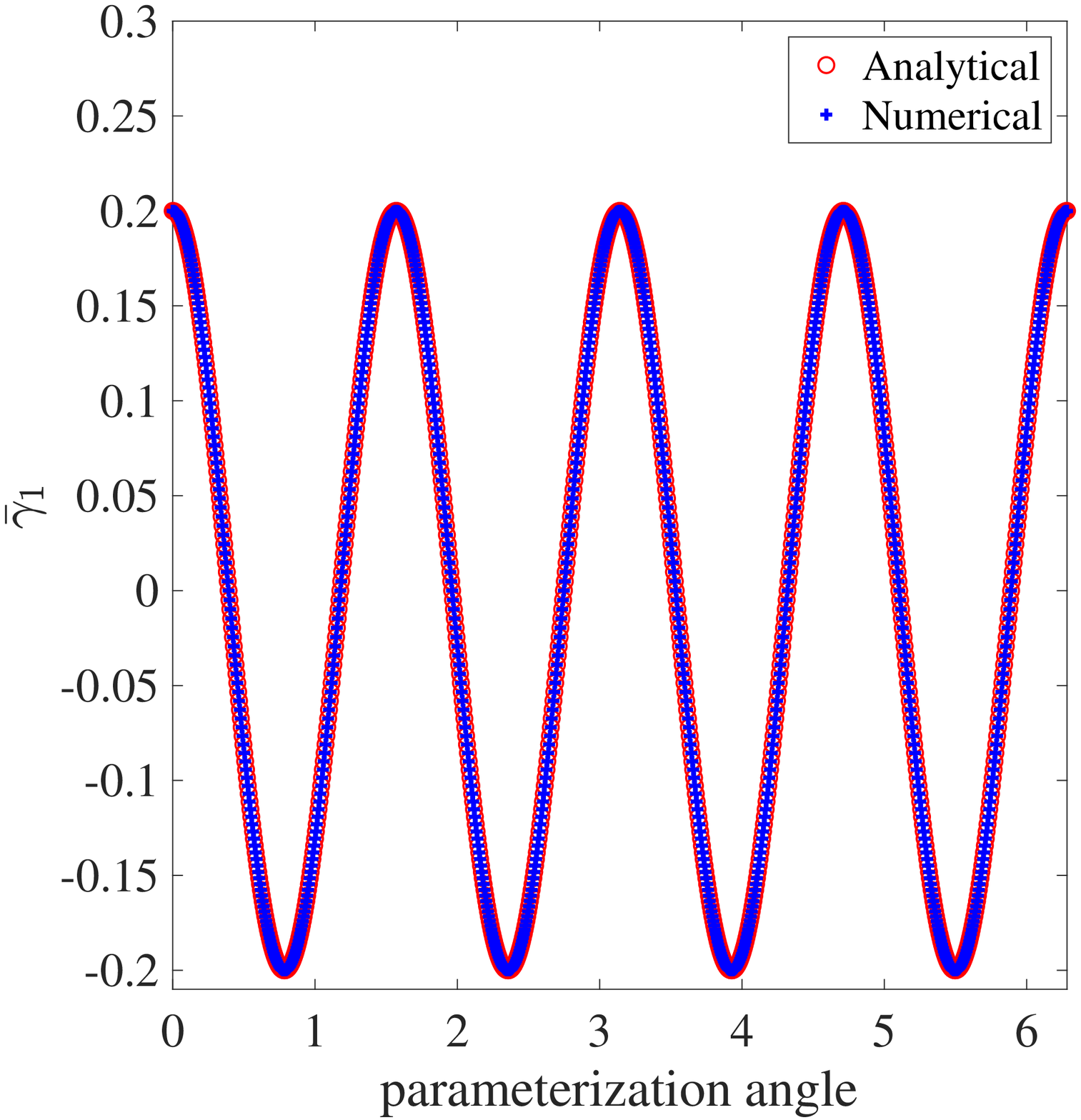}[a]
\includegraphics[scale=0.28]{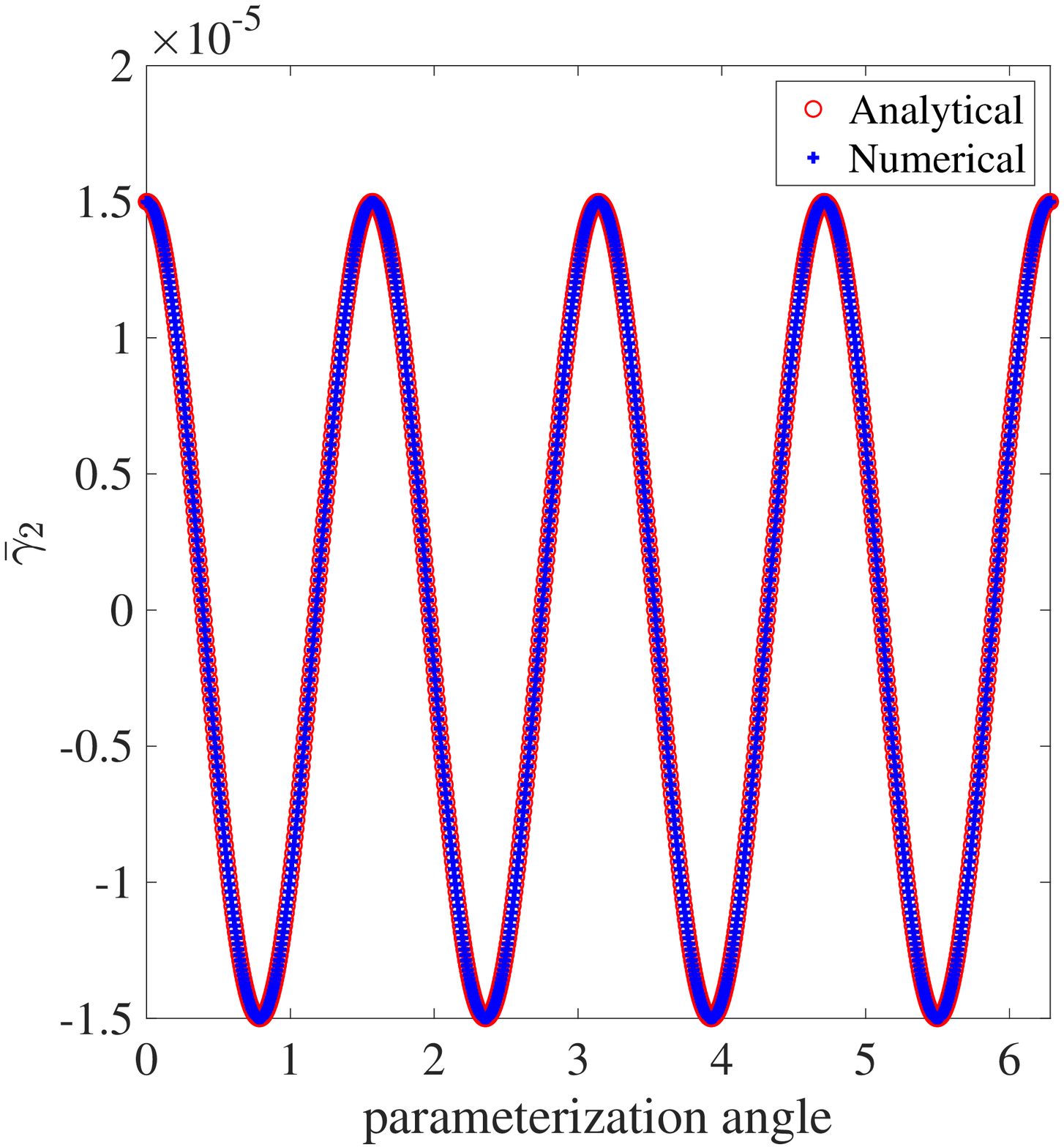}[b]
\caption{It shows the accuracy of the integral solver for a perturbed circle  $r(\theta,0)=1+0.1\times \cos(4\theta)$.}\label{Res:gamma}
\end{figure}

Finally, we check the accuracy of our scheme, which is basically the accuracy for solving the boundary integral equations \cref{integr12,integr22}. We assume analytical solutions $\displaystyle\bar\gamma_1=0.2\cos(4\theta)$ and $\displaystyle\bar\gamma_2=1.5\times 10^{-5}\cos(4\theta)$ to the integral equations with the shape to be $r(\theta,0)=1+0.1\times \cos(4\theta)$. We evaluate the left hand side of \cref{integr12,integr22}. Note here we are not able to find a closed form of the left hand side. These values are computed numerically by using the function {\it NIntegrate} in {\it Mathematica 11}. Applying these values as right hand side, we use our integral solver to compute $\bar\gamma_1$ and $\bar\gamma_2$. 
In this simulation, we use $N=1024$ points along the interface and no evolution is involved. In \cref{Res:gamma}, we compare the analytical solutions and our simulation results and the difference between these is around the machine epsilon. 

\subsection{Comparison with linear theory}
In this part, we compare the linear theory and nonlinear simulation.  We compute the shape factor numerically using $\displaystyle \left(\frac{\delta}{R}(t)\right)_{NL}=\max_{\theta}\left |{|\bar{\mathbf{x}}(\theta,t)|}/{{\bar R}_{eff}}-1\right |$,
where $\bar{\mathbf{x}}$ is the position vector measured from the centroid of the shape to the interface, $\displaystyle {\bar R}_{eff}=\sqrt{{\bar{A}}/{\pi}}$ is the effective radius of the viscous fluid in the rescaled frame and $\bar{A}$ is the constant area enclosed by the interface. ƒs

\begin{figure}[tbhp]
\centering
\includegraphics[scale=0.28]{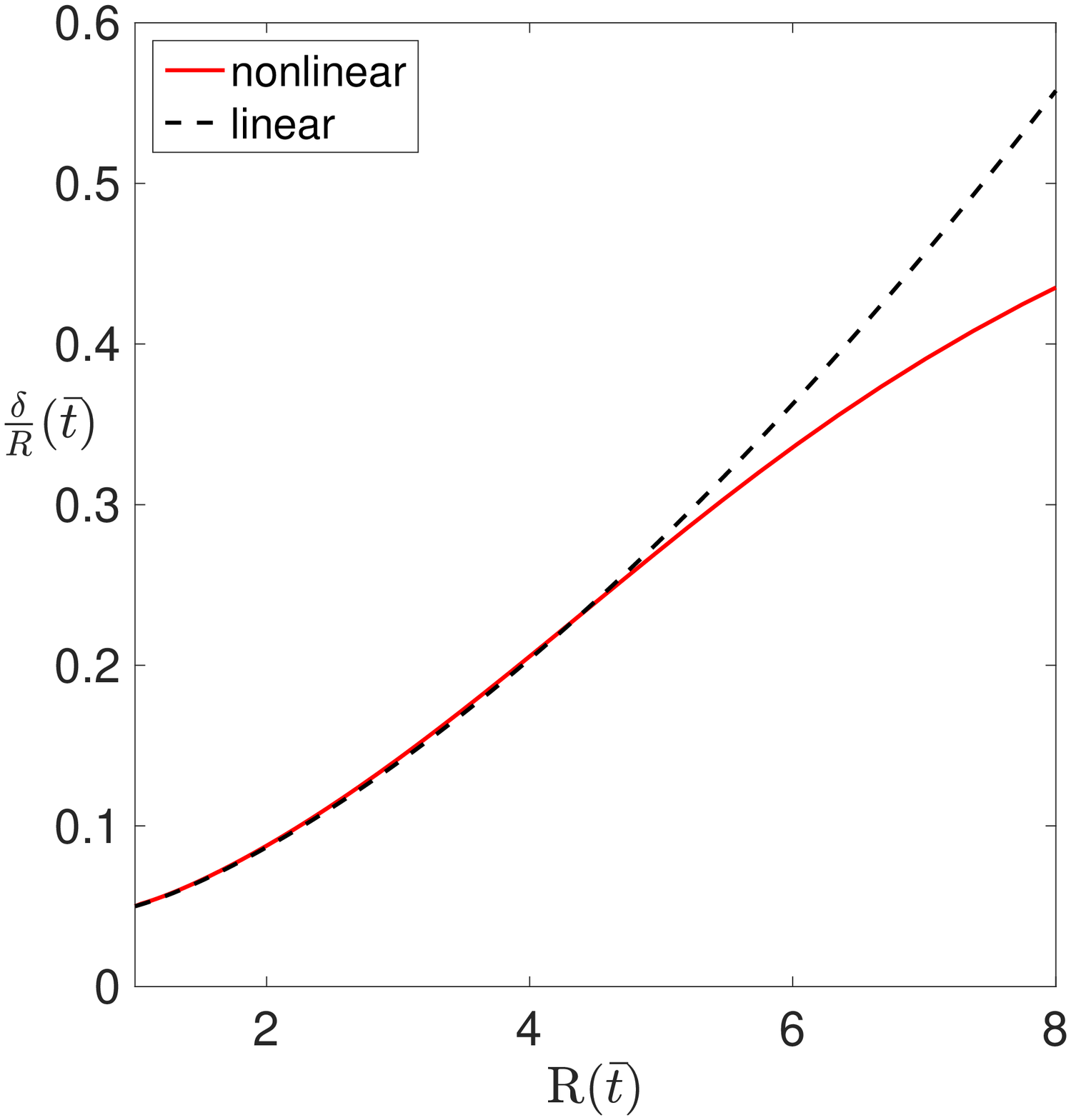}[a]
\includegraphics[scale=0.28]{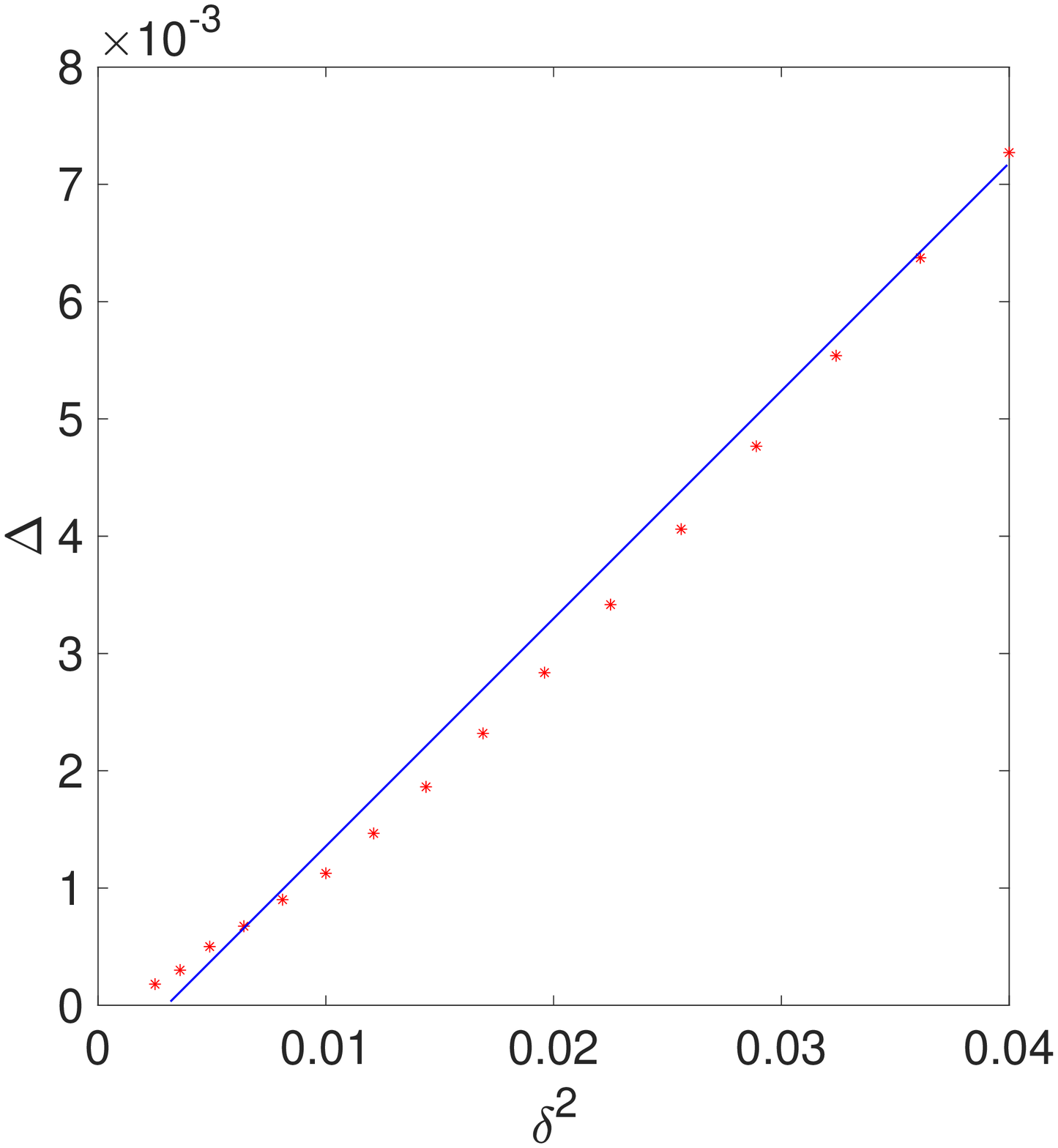}[b]
\caption{Comparison with linear theory. [a] shows the linear comparison between linear theory and numerical results. [b] shows the difference of the shape factor between
  linear and nonlinear solutions at various initial perturbations. }
  \label{Res:fig2}
\end{figure}

We set $N=4096$ points along the
interface and the time step $\Delta \bar{t}=1\times 10^{-3}$.  The initial interfacial condition is taken as $r(\theta,0)=1+0.05\times \cos(4\theta)$. All other parameters are the same as utilized in \cref{Res:fig1,Res:gamma}. We denote $\displaystyle (\frac{\delta}{R})_{NL}$  and $\displaystyle (\frac{\delta}{R})_{Lin}$
as the the nonlinear and linear shape factor
respectively. From the linear analysis, we have  
\begin{multline}
(\frac{\delta}{R})_{Lin}=(\frac{\delta}{R})_{0}R^{\frac{(k_{eo_2}k_{h_1}-k_{eo_1}k_{h_2})}{(k_{eo_1}+k_{eo_2})^2-(k_{e_1}+k_{e_2})(k_{h_1}+k_{h_2})}\frac{2nI}{J}+(n\frac{k_{eo_1}^2-k_{eo_2}^2-(k_{e_1}+k_{e_2})(k_{h_1}-k_{h_2})}{(k_{eo_1}+k_{eo_2})^2-(k_{e_1}+k_{e_2})(k_{h_1}+k_{h_2})}-2)}\\
\exp{[\frac{(k_{eo_2}^2k_{h_1}+(k_{eo_1}^2-(k_{e_1}+k_{e_2})k_{h_1})k_{h_2})}{(k_{eo_1}+k_{eo_2})^2-(k_{e_1}+k_{e_2})(k_{h_1}+k_{h_2})}\frac{n(n^2-1)\tau}{J}(R^{-1}-1)]}.
\end{multline}
 As shown in \cref{Res:fig2}[a], we get a good agreement
between the linear and nonlinear results at early times. As time progresses, the interface produce long fingers and nonlinear effects make the numerical curve grows less than the linear theory curve. We next vary the initial perturbation of the interface, $r=1+\delta\cos(4\theta)$, without changing other parameters, where $\delta$ differs from $0.05$ to $0.2$. We run the scheme till $\bar{T}=0.1$. We calculate the difference between the linear and nonlinear
results and denote the
difference as $\displaystyle
\Delta=|(\frac{b}{R})_{NL}-(\frac{b}{R})_{Lin}|$. It is
expected that the difference should be at the order of $\delta^2$,
as shown in \cref{Res:fig2}[b].

\subsection{Simulations under constant flux $J=1$}
\begin{figure}[tbhp]
\includegraphics[scale=0.14]{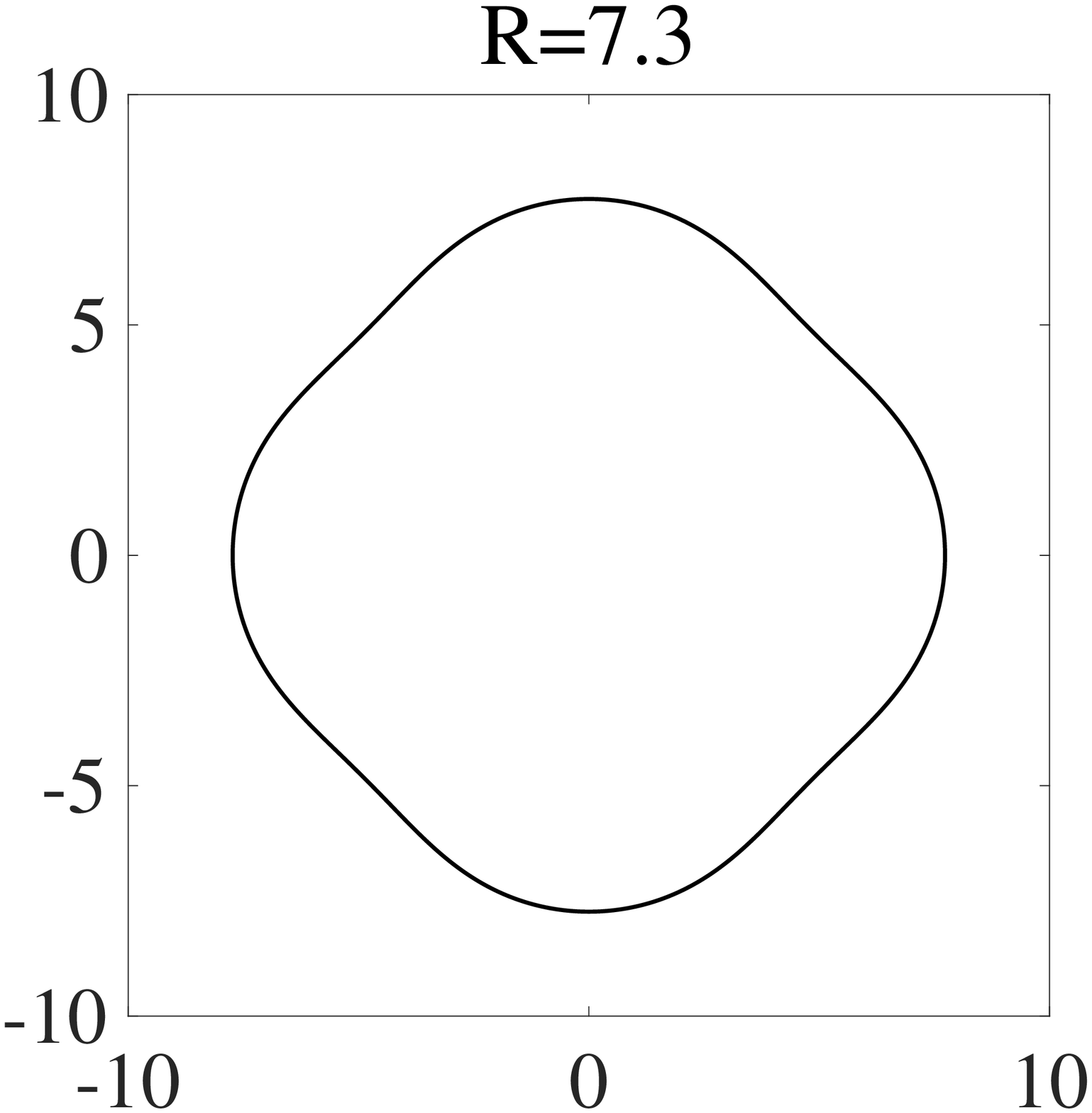}
\includegraphics[scale=0.14]{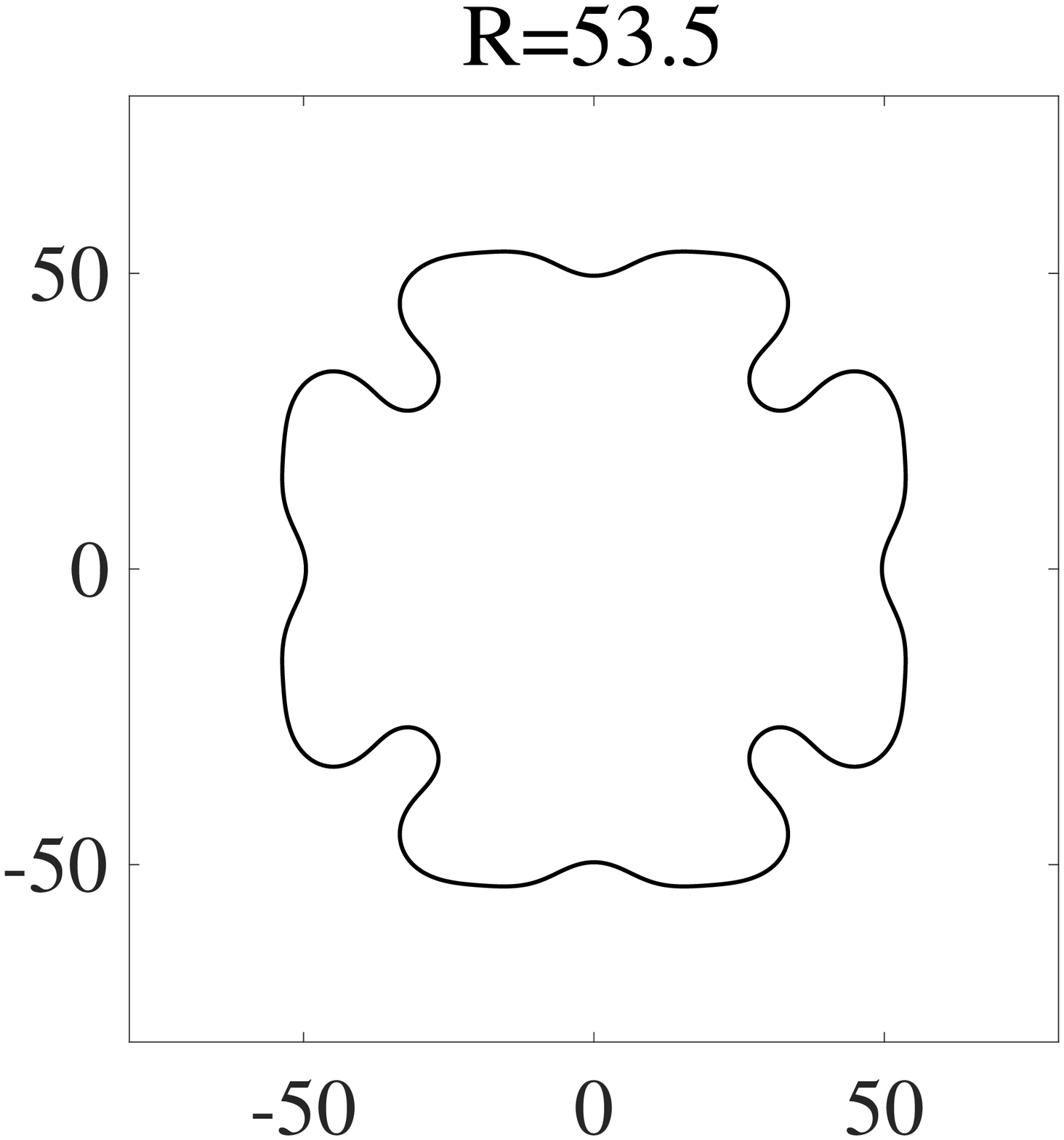}
\includegraphics[scale=0.14]{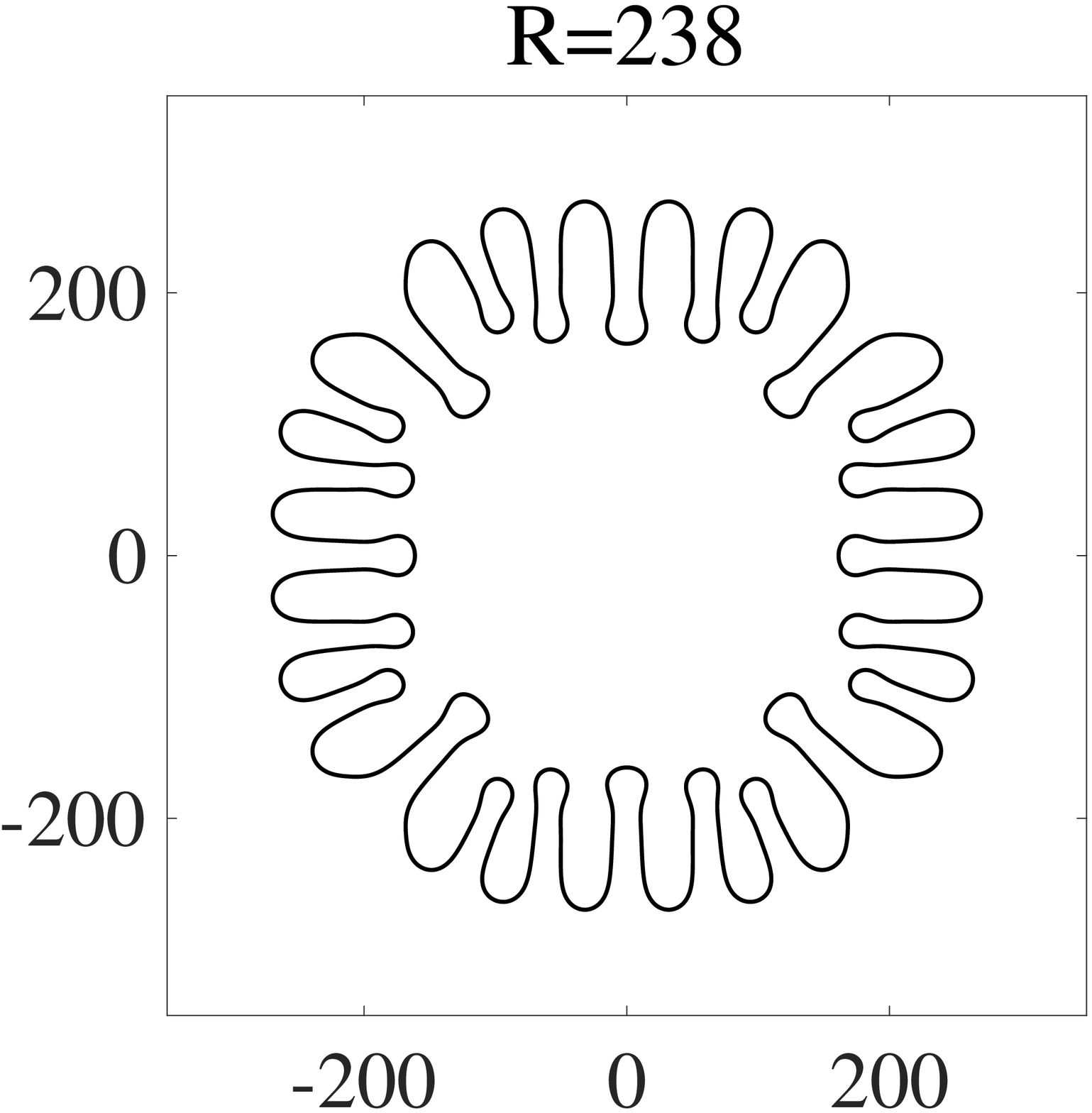}
\includegraphics[scale=0.14]{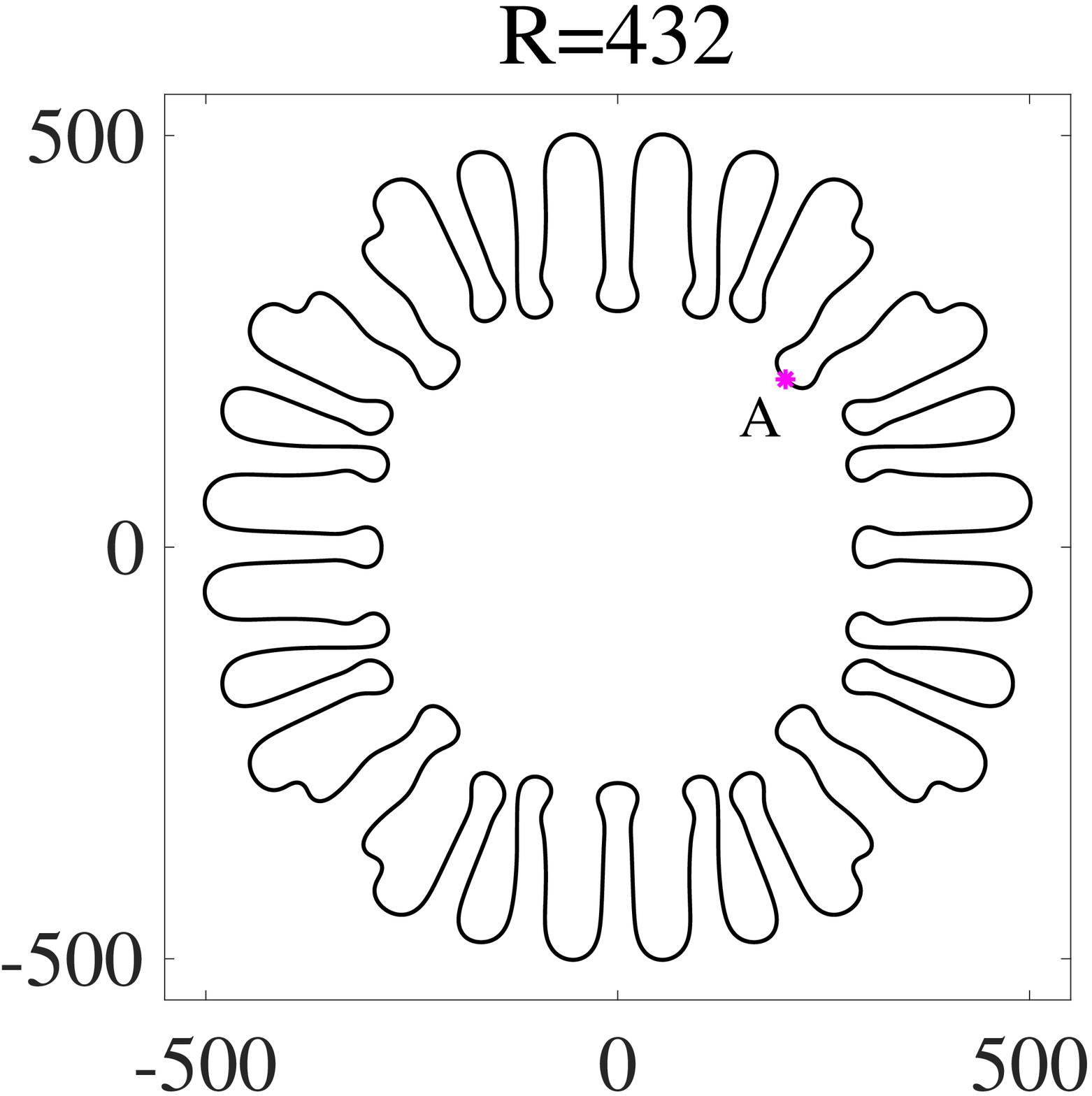}\\
\includegraphics[scale=0.14]{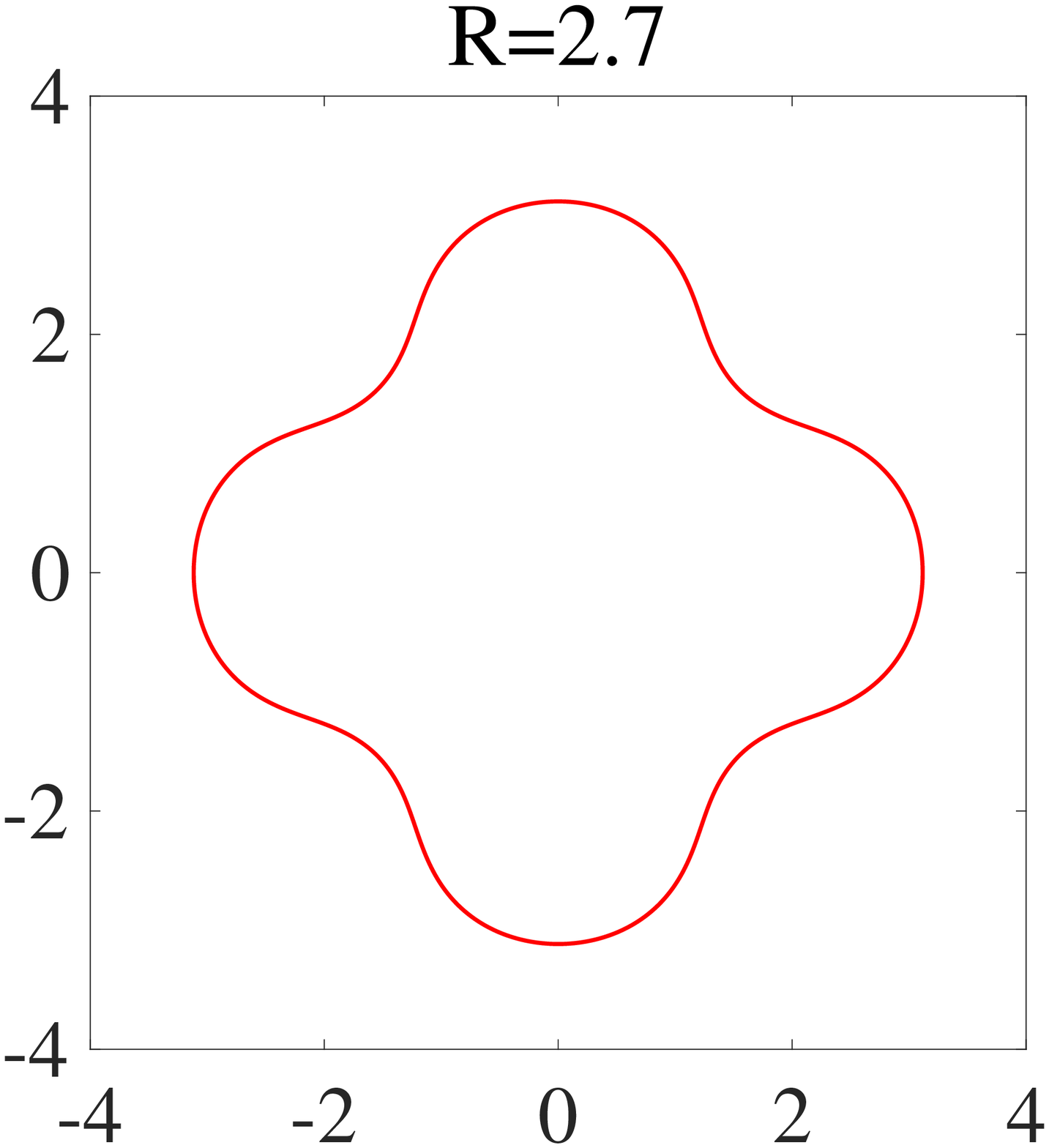}
\includegraphics[scale=0.14]{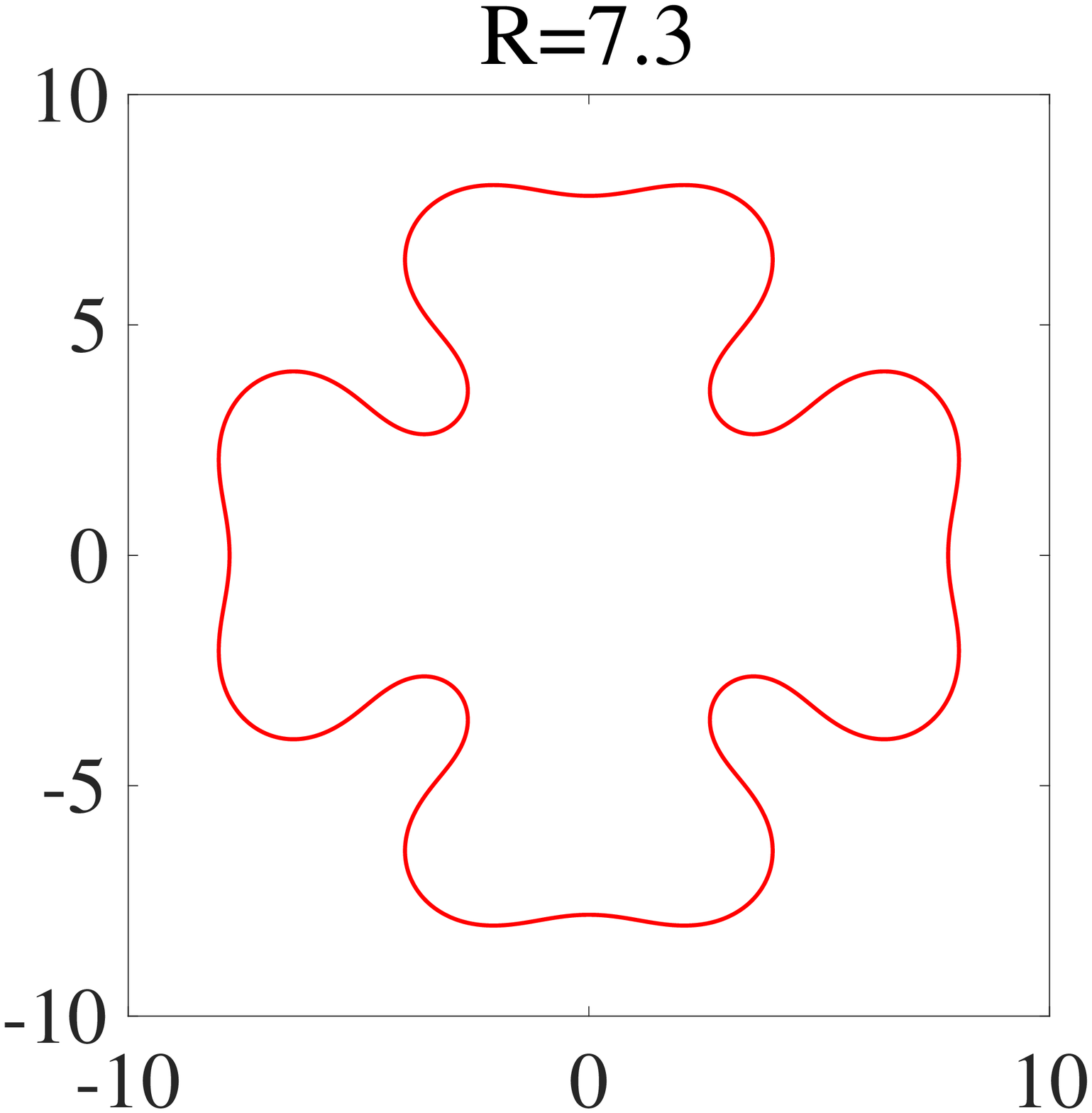}
\includegraphics[scale=0.14]{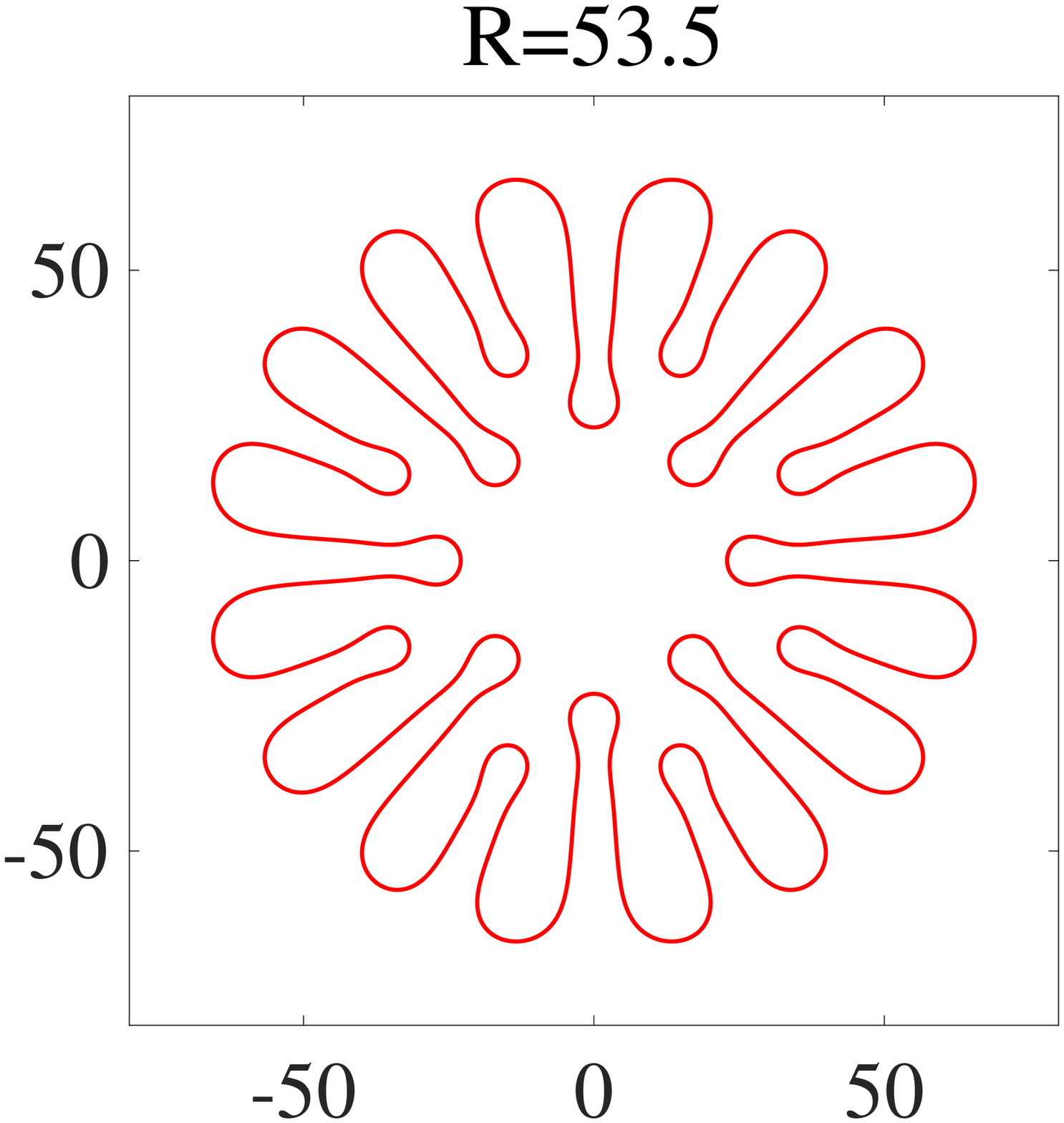}
\includegraphics[scale=0.14]{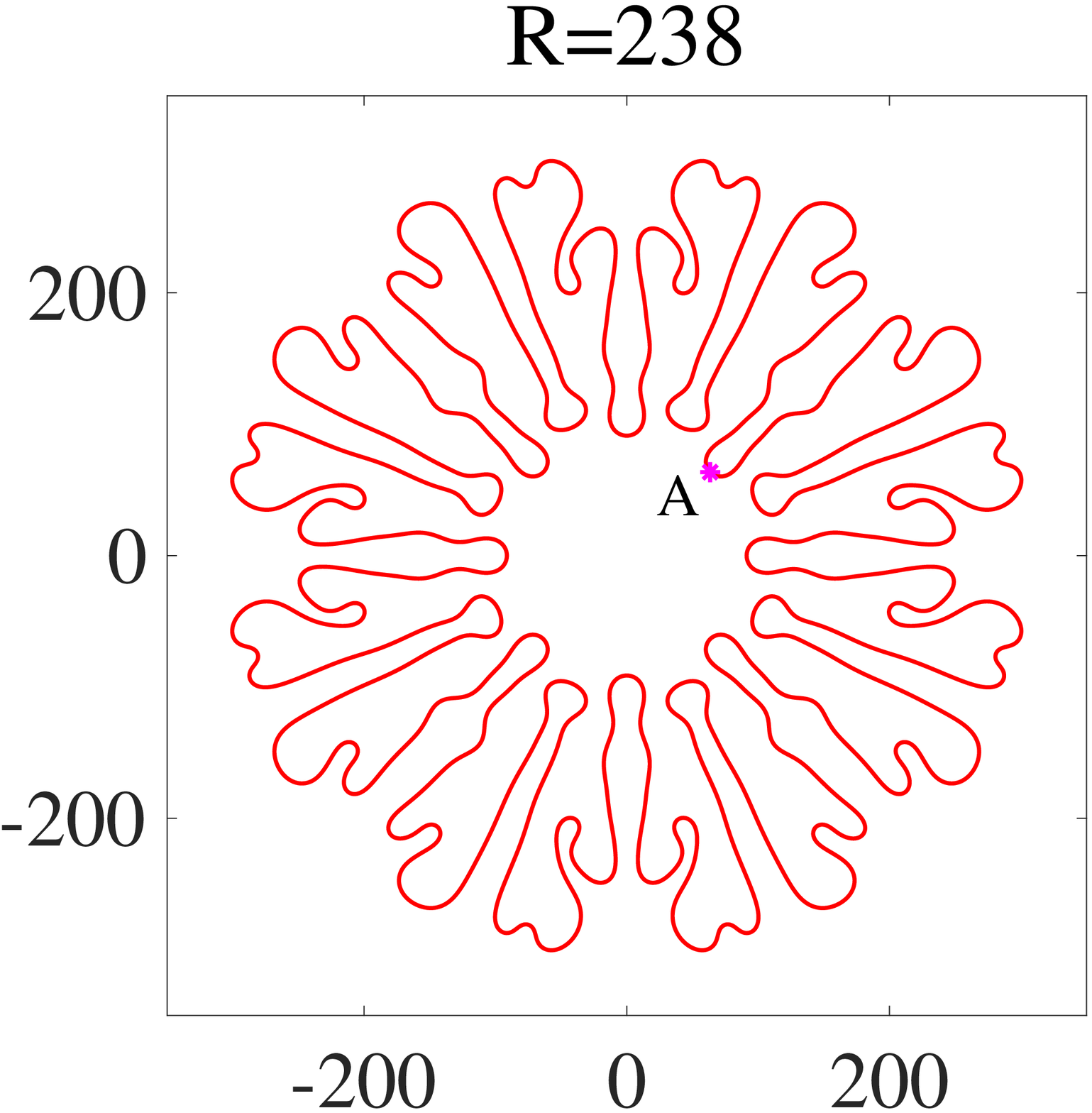}\\
\includegraphics[scale=0.14]{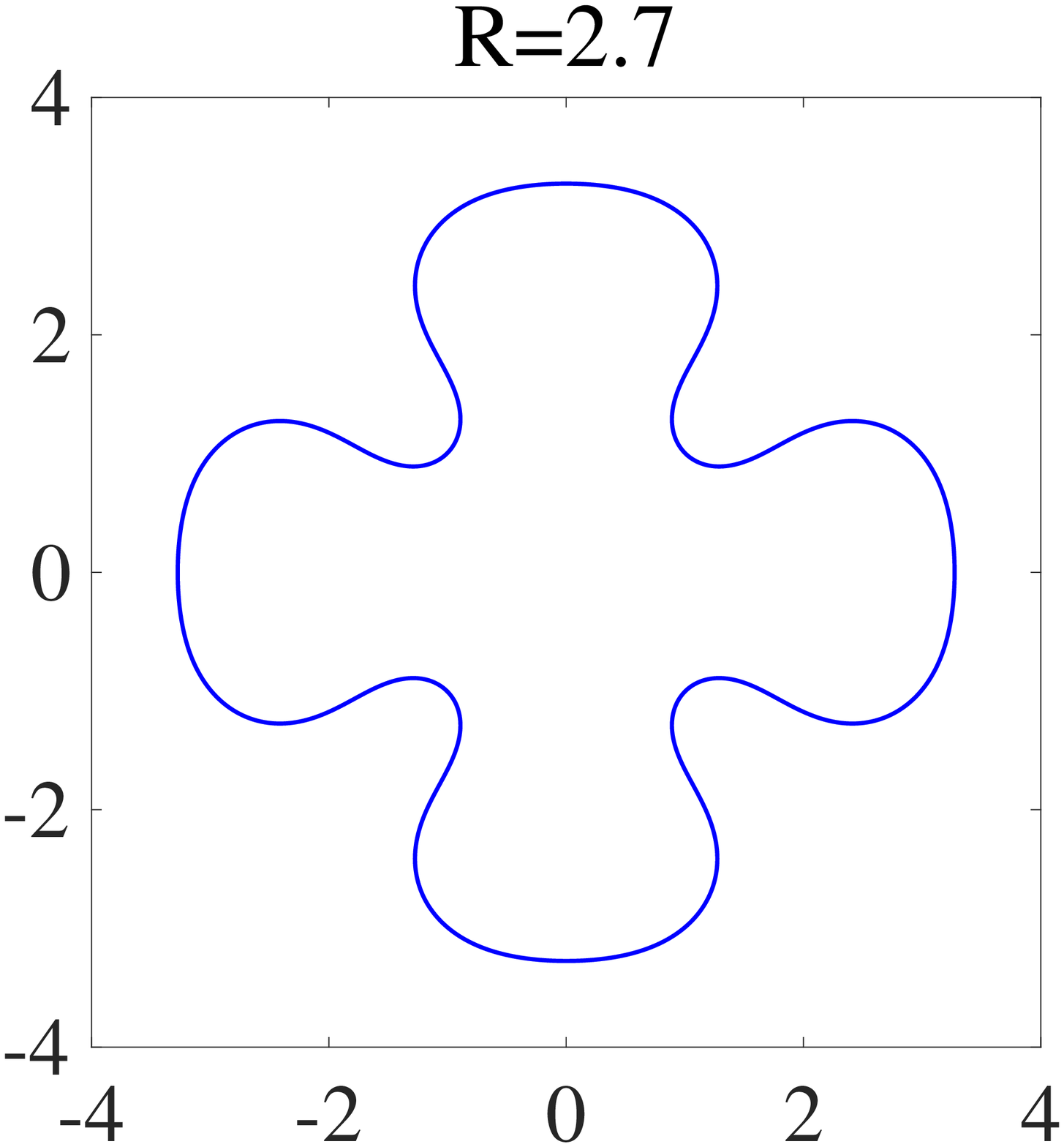}
\includegraphics[scale=0.14]{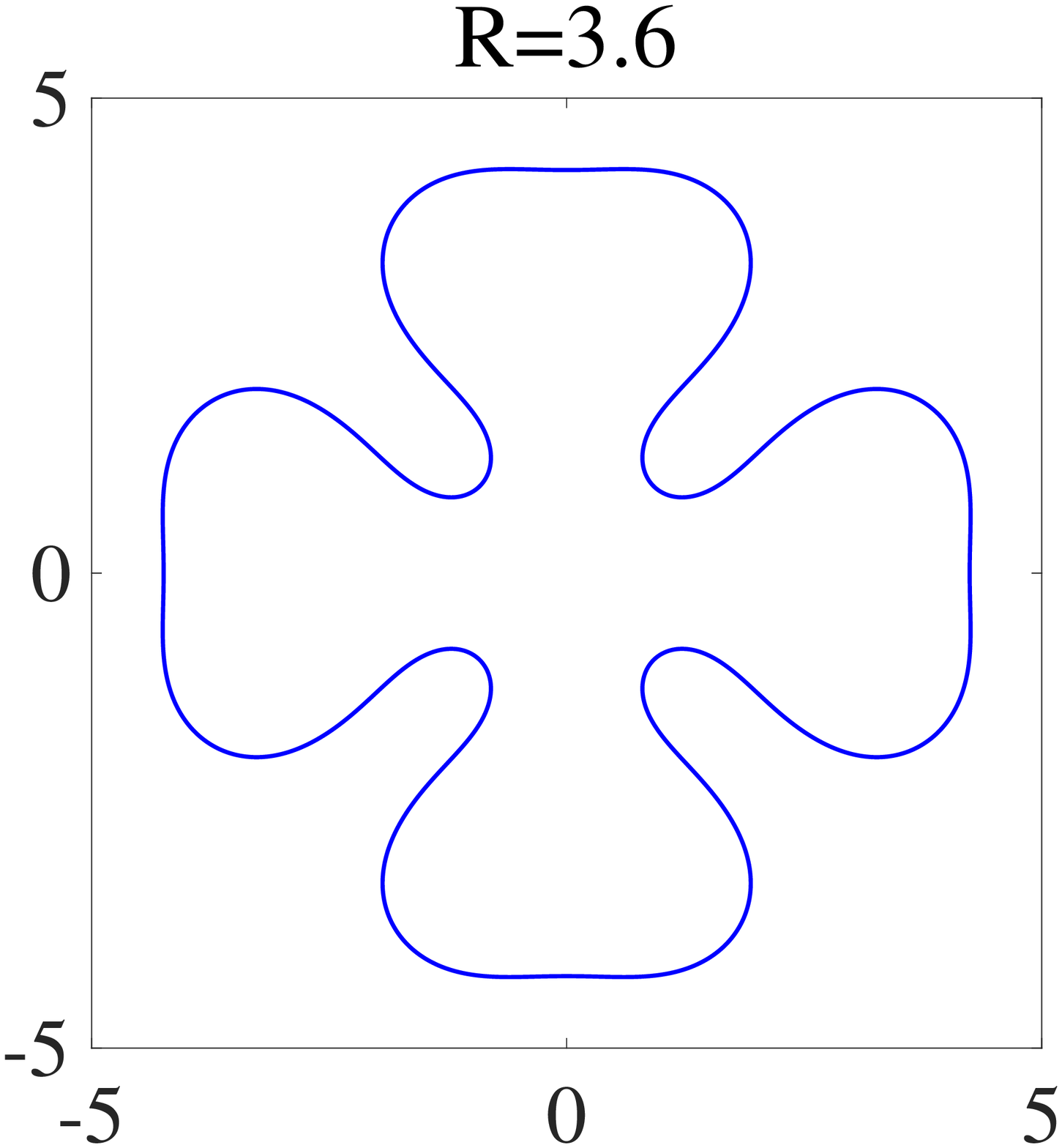}
\includegraphics[scale=0.14]{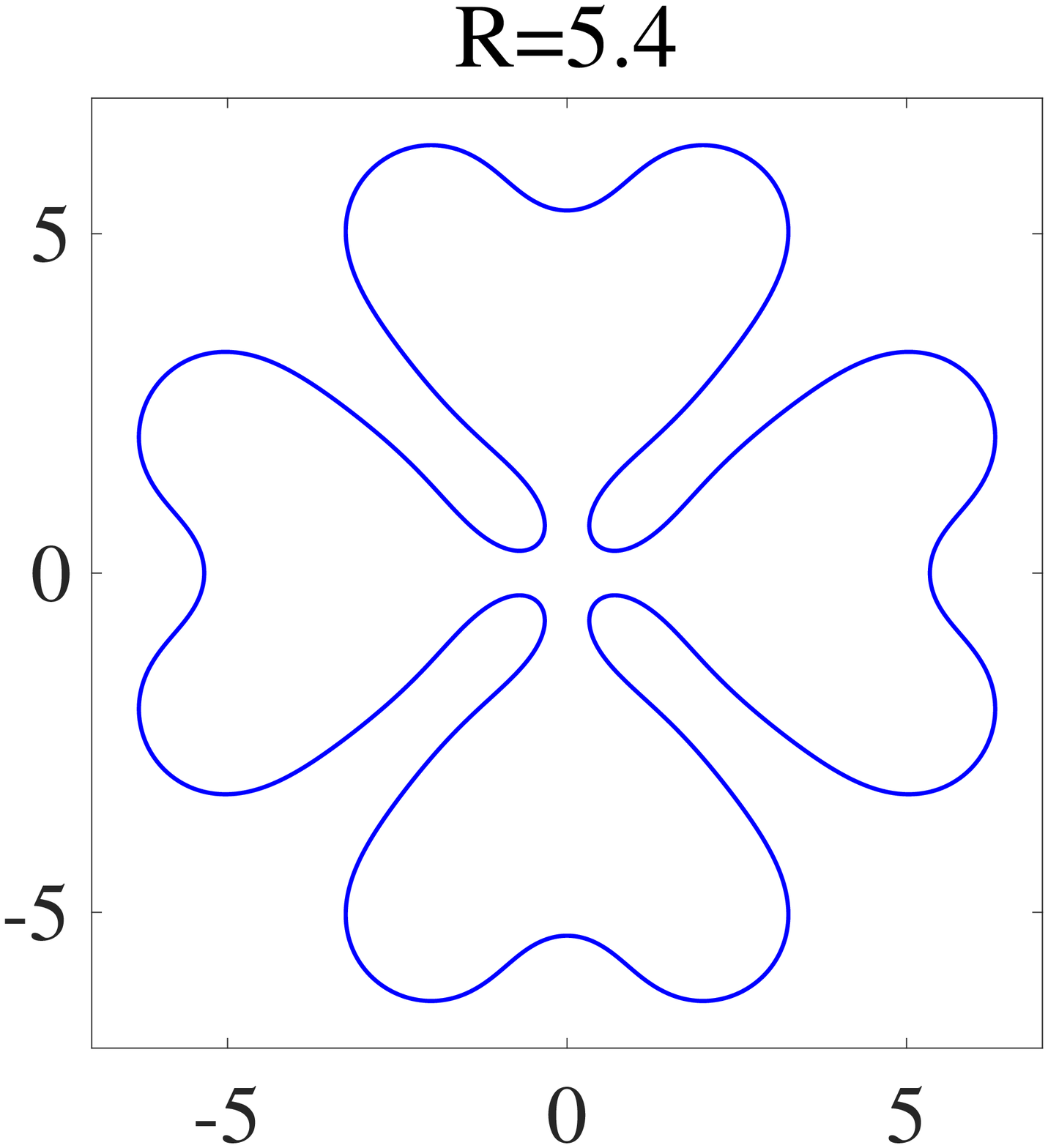}
\includegraphics[scale=0.14]{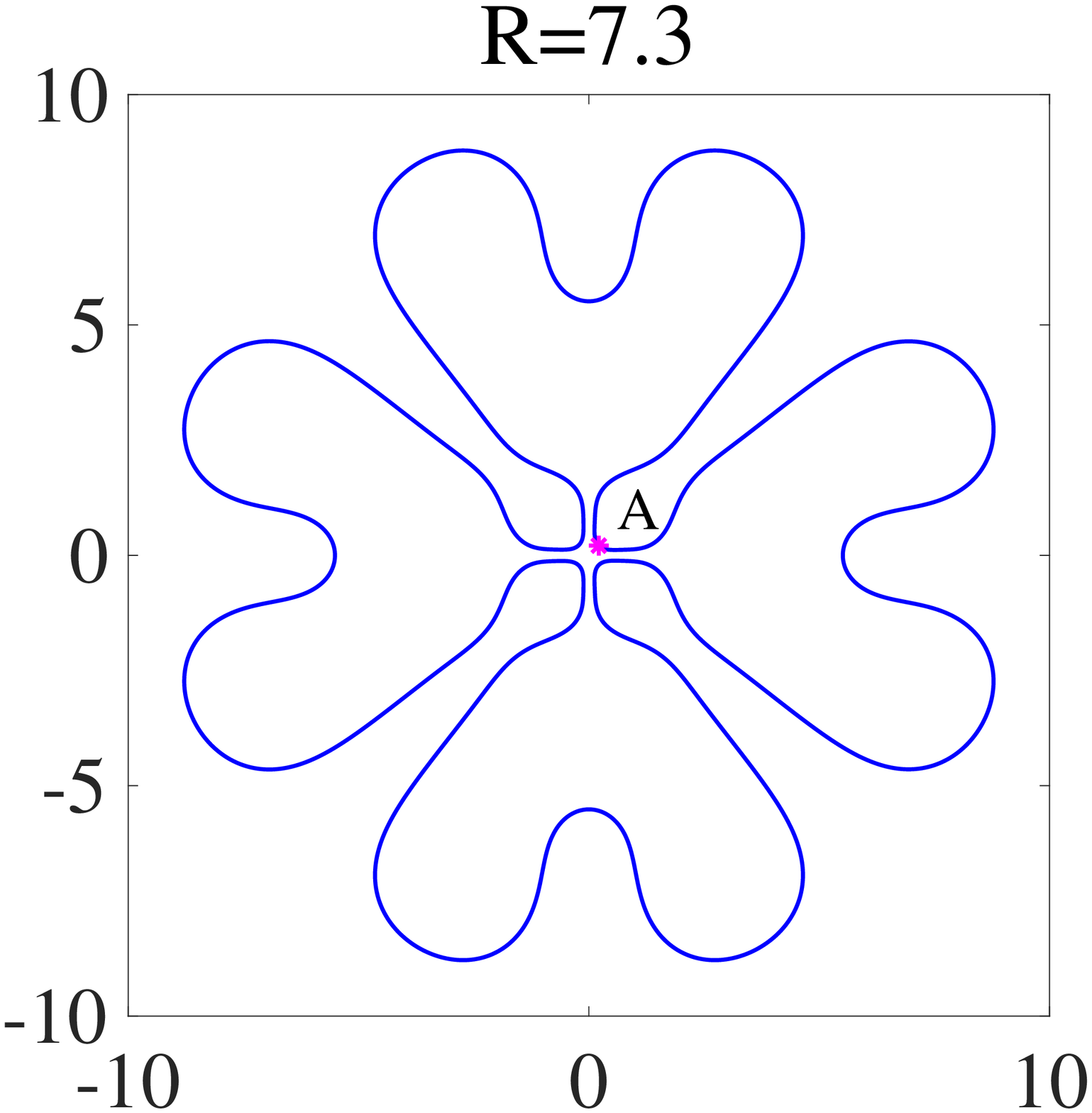}[a]\\
\includegraphics[scale=0.28]{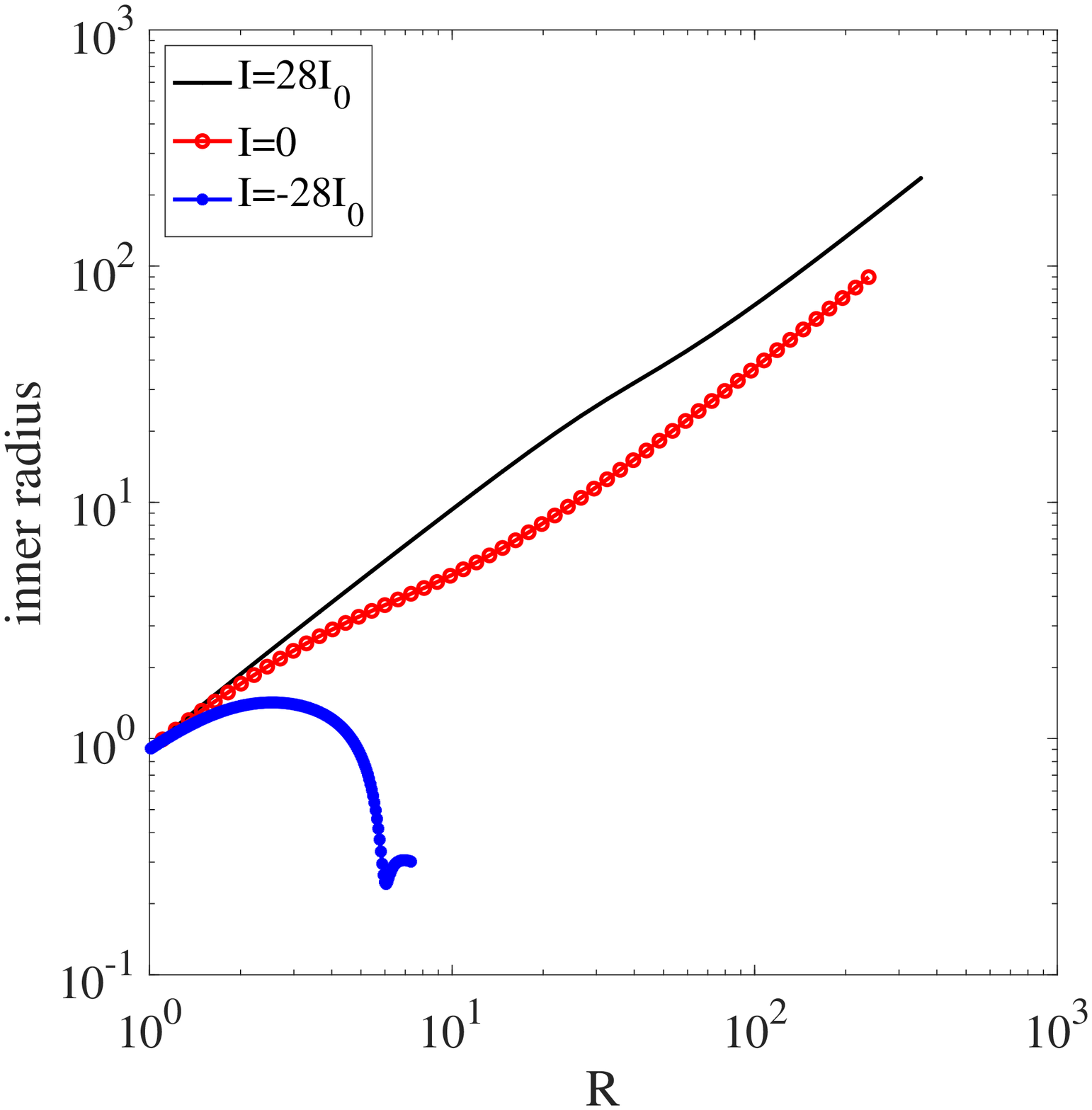}[b]
\includegraphics[scale=0.28]{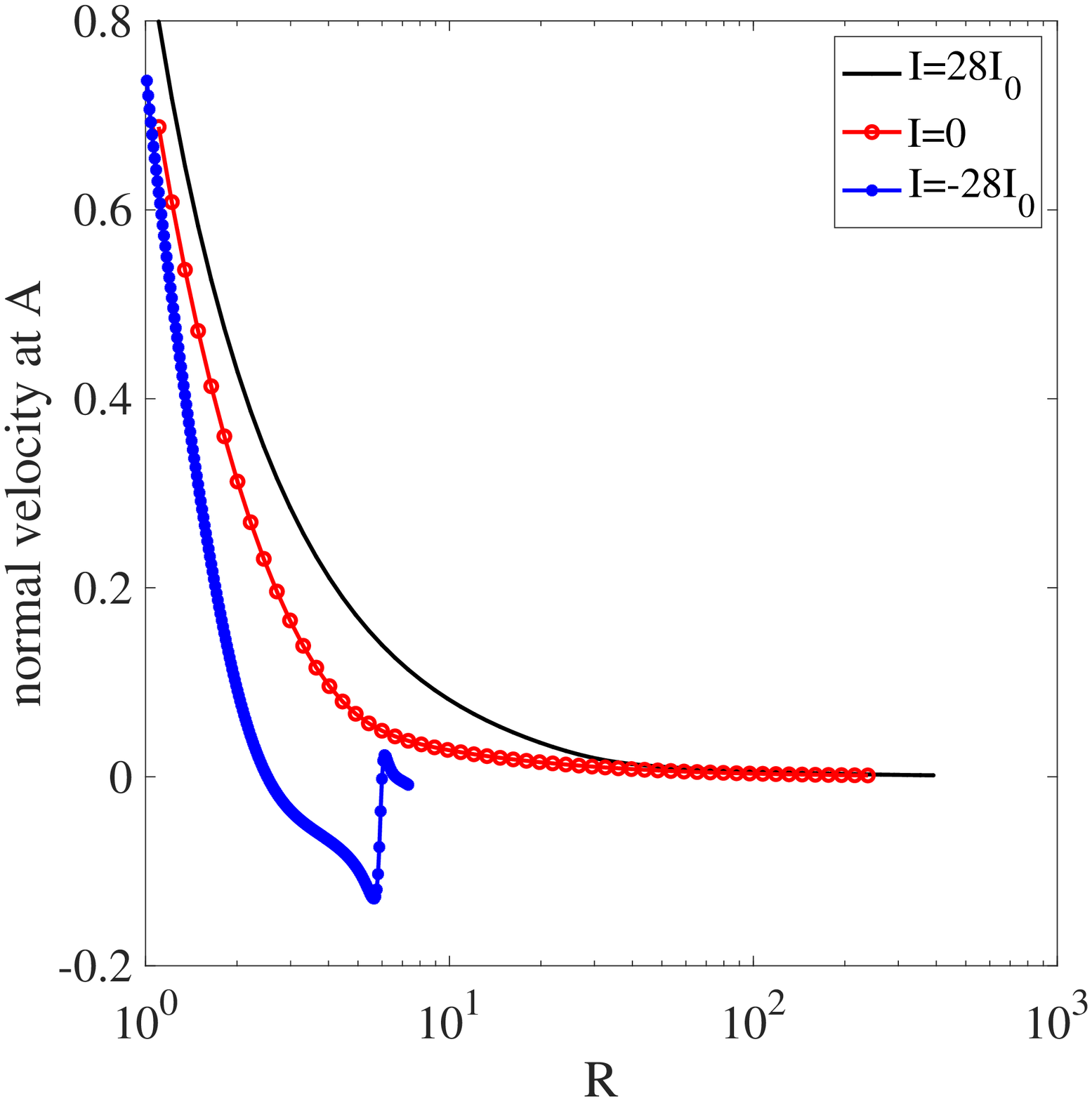}[c]
\caption{The interface dynamics under constant flux $J=1$, surface tension $\tau=2.16\times 10^{-2}$ and various currents $I$. In [a], interface morphologies under $I=28I_0$  are shown in the first row (black); morphologies under $I=0$ are shown in the second row (red); and morphologies under $I=-28I_0$  are shown in the third row (blue). In all cases, the interfaces are unstable and develop multiple fingers. [b] shows the inscribed circle radius as a function of $R$. [c] shows the normal velocity of point A on the interface}\label{Res:fig4}
\end{figure}

First we study the interface dynamics under constant flux $J=1$ and surface tension $\tau=2.16\times 10^{-2}$. Three different constant currents are used and the morphologies of the interface are shown in \cref{Res:fig4}[a]. When $I=0$ (second row, red patterns) is applied, which indicates that the system is pure hydrodynamic, the interface experiences the well-know fingering instabilities and repeated tip-splitting process. Comparing the red patterns with the black morphologies at $R=7.3$, we find that a positive current $I=28I_0$ (first row, black patterns) exhibits an effect of stabilization, where the interface shows a compact shape.  Although the interface under the positive current develops multiple fingers at $R=53.5$ and $R=238$, the interface does not experience repeated tip-splitting. Also, the amplitudes of the finger for $I=0$ case are larger than the lengths of the fingers found for $I>0$ when compared at similar interfacial sizes.
 On the other hand, a negative current $I=-28I_0$ (third row, blue patterns) promotes interface instabilities, where the interface produces wide fingers and thin tails connecting fingers.  

\begin{figure}[tbhp]
\includegraphics[scale=0.14]{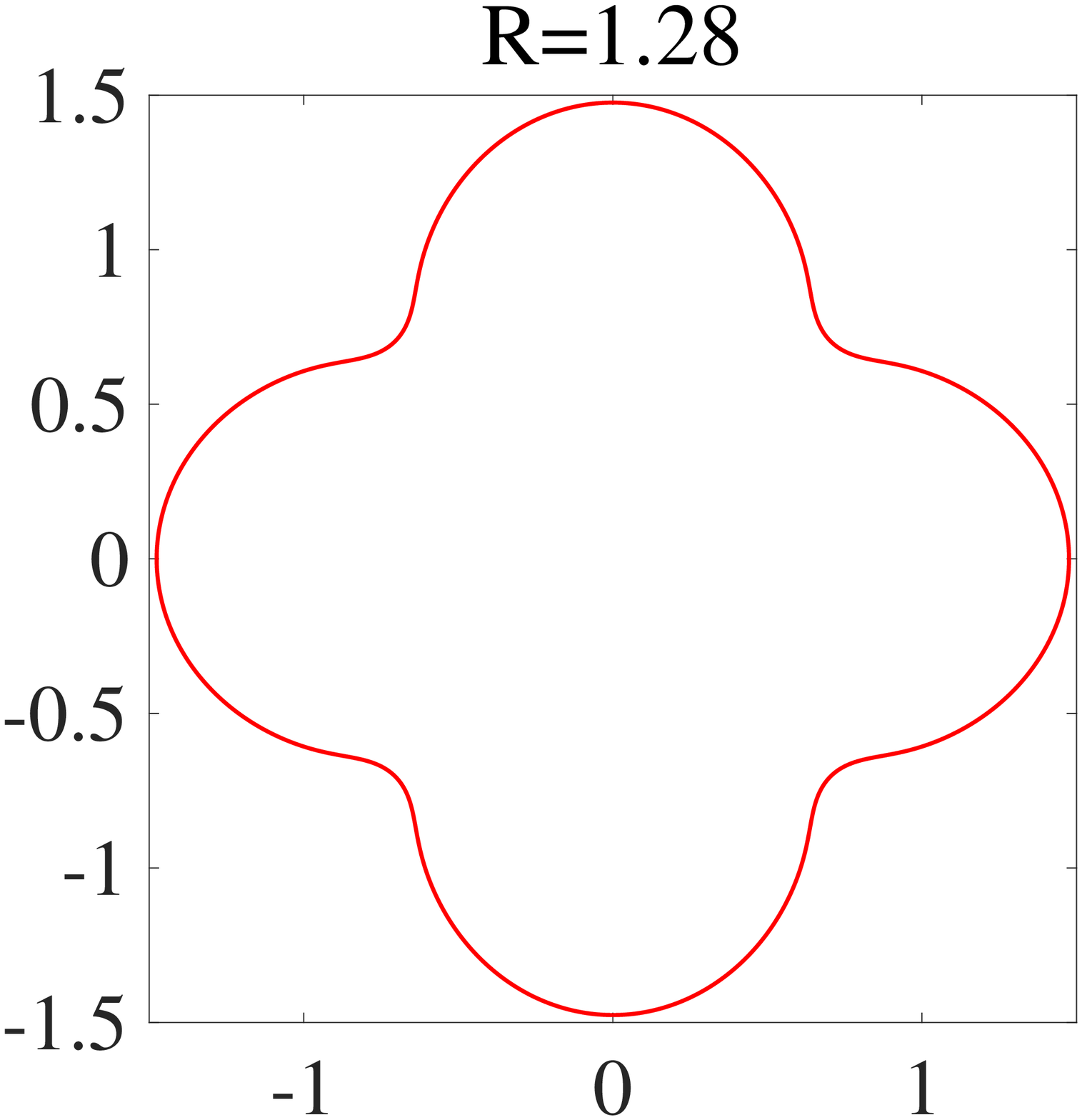}
\includegraphics[scale=0.14]{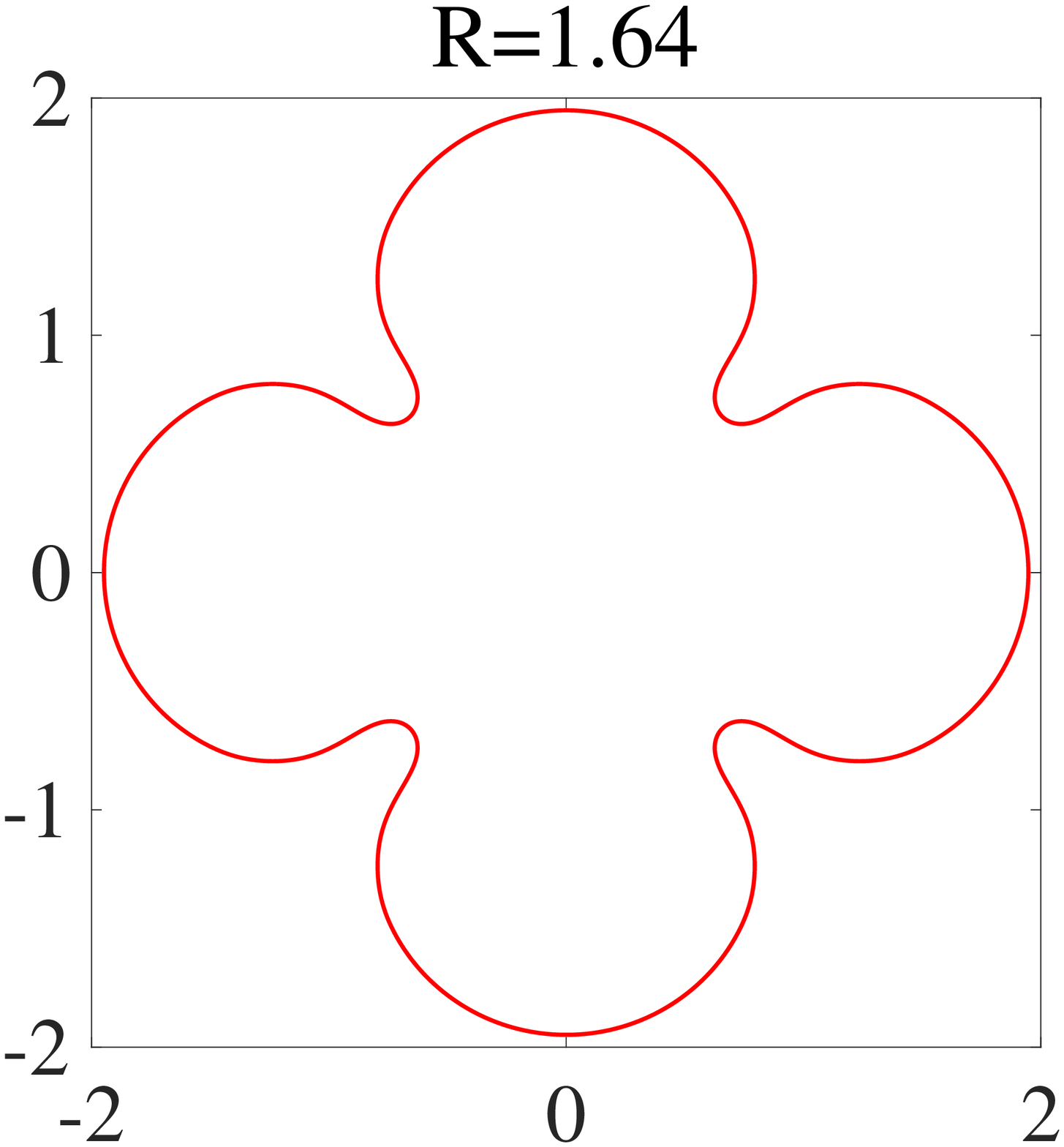}
\includegraphics[scale=0.14]{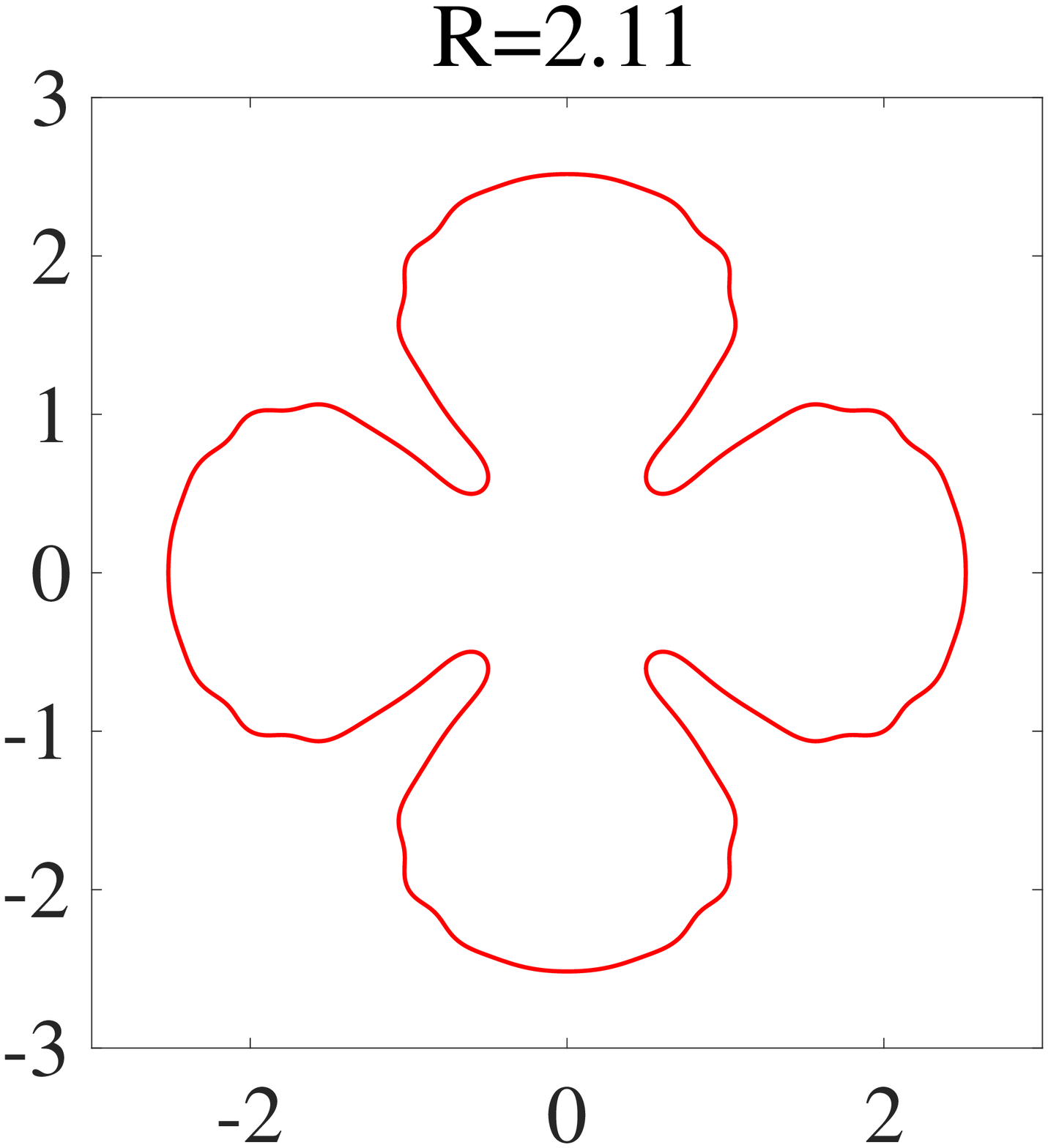}
\includegraphics[scale=0.14]{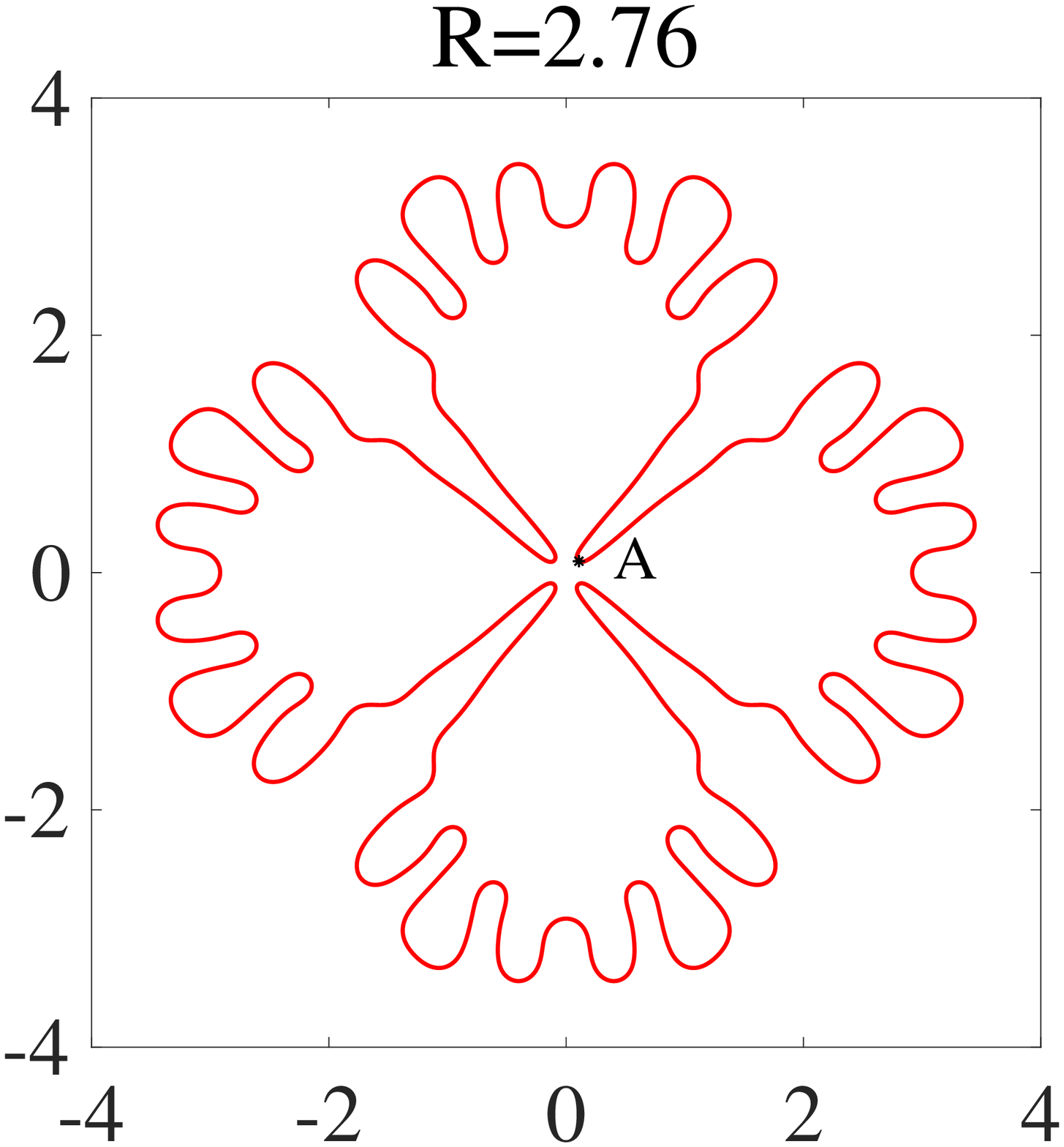}[a]\\
\includegraphics[scale=0.28]{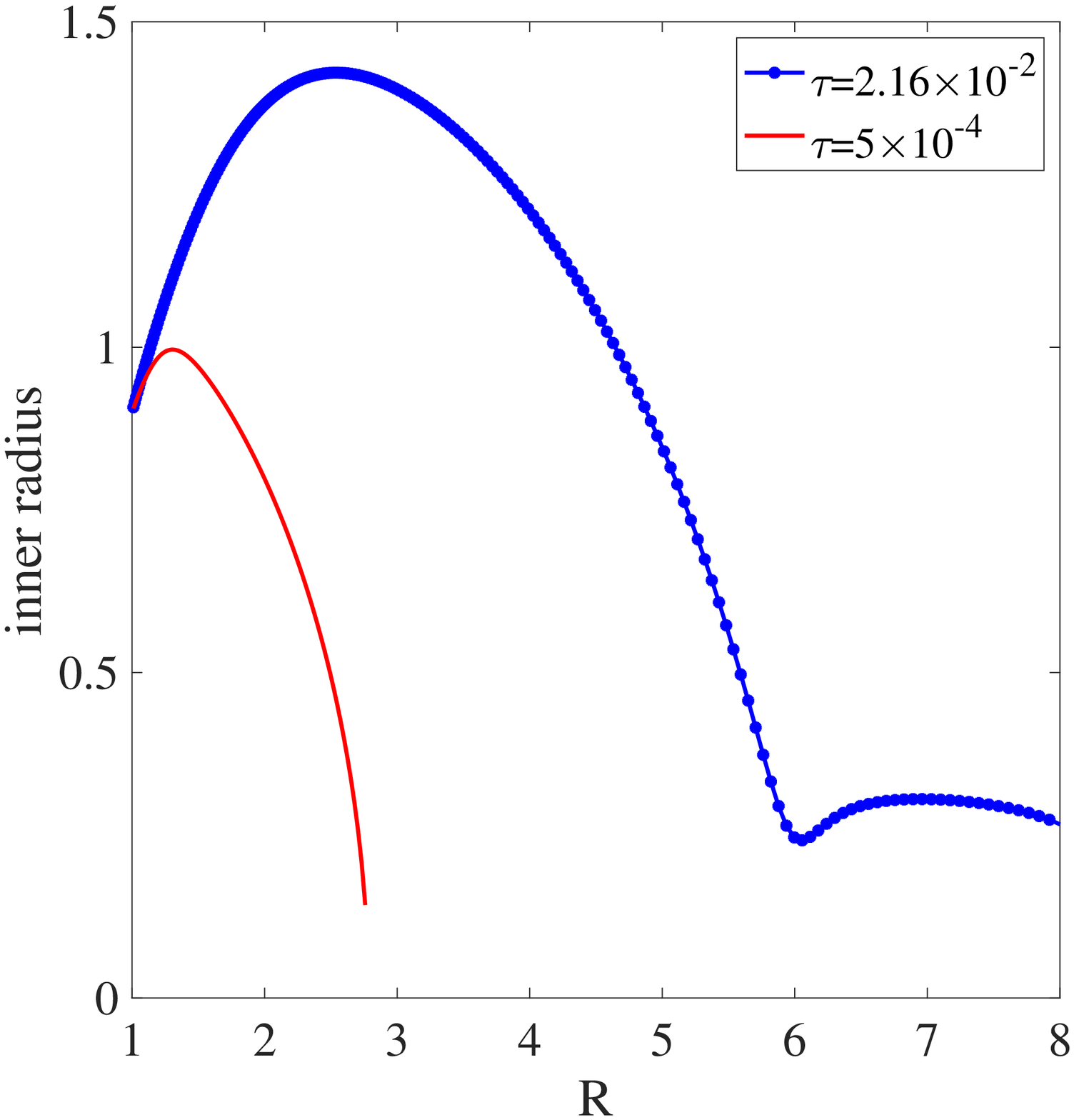}[b]
\includegraphics[scale=0.28]{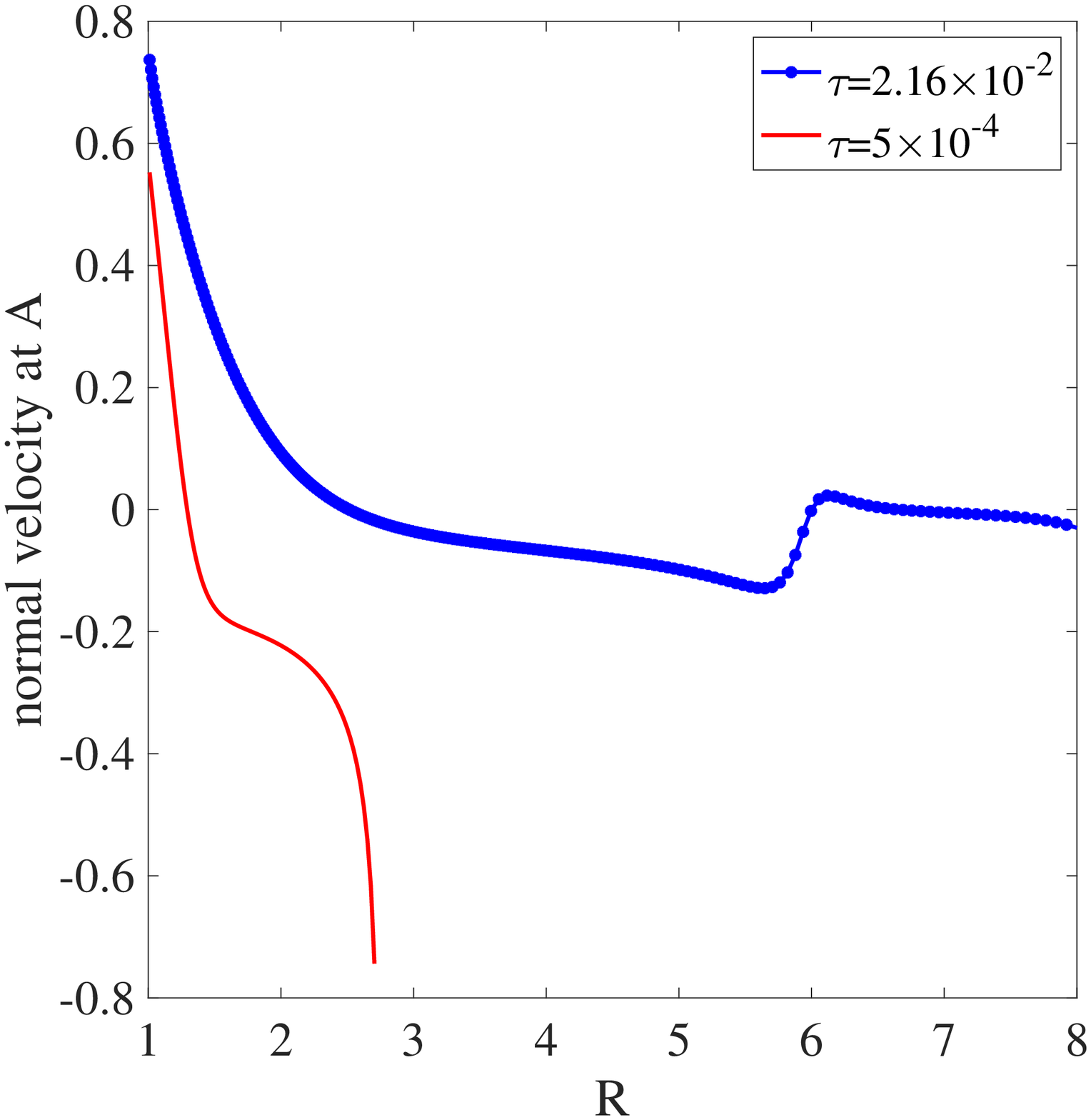}[c]
\caption{The interface dynamics under constant flux $J=1$, constant current $I=-28I_0$,  and surface tension  $\tau=5\times 10^{-4}$. [a] shows the morphologies of the interface.  [b] shows the inscribed circle radius as a function of $R$. [c] shows the normal velocity of point A.}\label{Res:fig6}
\end{figure}

In \cref{Res:fig4}[b], we consider the inscribed circle radius of the interface, which is the smallest distance from the interface to origin. The inner radius grows as the interface expands radially outward for cases $I=28I_0$ and $I=0$. In addition, at later times, the inner radius is proportional to $R$, with coefficient $0.65$ and $0.38$ for $I=28I_0$ and $I=0$, respectively.
For  $I=-28I_0$, the distance increases at early times as the interface expands. However, due to the unstable effects induced by the negative current, the interface develops long fingers, and point A, as indicated in the pattern morphologies shown in \cref{Res:fig4}[a], moves inward resulting a decrease in the inner radius. After that, point A tends to moves outwards and becomes almost stagnant. This phenomena can be explained by investigating the normal velocity of point A (see \cref{Res:fig4}[c]). The normal velocity drops dramatically at early times. Later the velocity behaves like $R^{-b}$, where $b=0.947$ and $0.955$ for  $I=28I_0$ and $I=0$, respectively. Alternatively, the velocity becomes differently for $I=-28I_0$ which can be observed more clearly in \cref{Res:fig6}. % Meanwhile, the interface forms thin tail regions. 

In \cref{Res:fig6}, we study the interface evolution under a smaller surface tension than the value utilized previously in \cref{Res:fig4}. We apply a constant flux $J=1$, constant current $I=-28I_0$ and surface tension $\tau=5\times 10^{-4}$. Comparing the morphologies here with those in the third row of \cref{Res:fig4}[a], we find that smaller surface tension promotes the interfacial instability. The interface under $\tau=5\times 10^{-4}$ exhibits more fingers and larger finger size. Also, the interface under smaller surface tension does not develop thin tail regions, which reveals that the tail region is a balance between the destabilizing (flux and current) and stabilizing (surface tension) effects.  Again, we analyze the inscribed circle radius (shown in \cref{Res:fig6}[b]). The inner radius in large surface tension case tends to be a constant (stagnating point) after a period of decrease. On the other hand, the small surface tension case tends to drop quickly to zero. Investigating the normal velocity at point A, we find the velocity under $\tau=2.16\times 10^{-2}$ becomes negative indicating that A moves inward. Then the velocity changes to positive and decays gradually around zero. But the velocity under $\tau=5\times 10^{-4}$ becomes begative and increases in magnitude, indicating that point A moves inward to the origin faster as time elapses. Combined these findings, we conjecture that the interface may touch the origin at a finite time. 

\subsection{Simulations under zero flux $J=0$}
\begin{figure}[tbhp]
\includegraphics[scale=0.3]{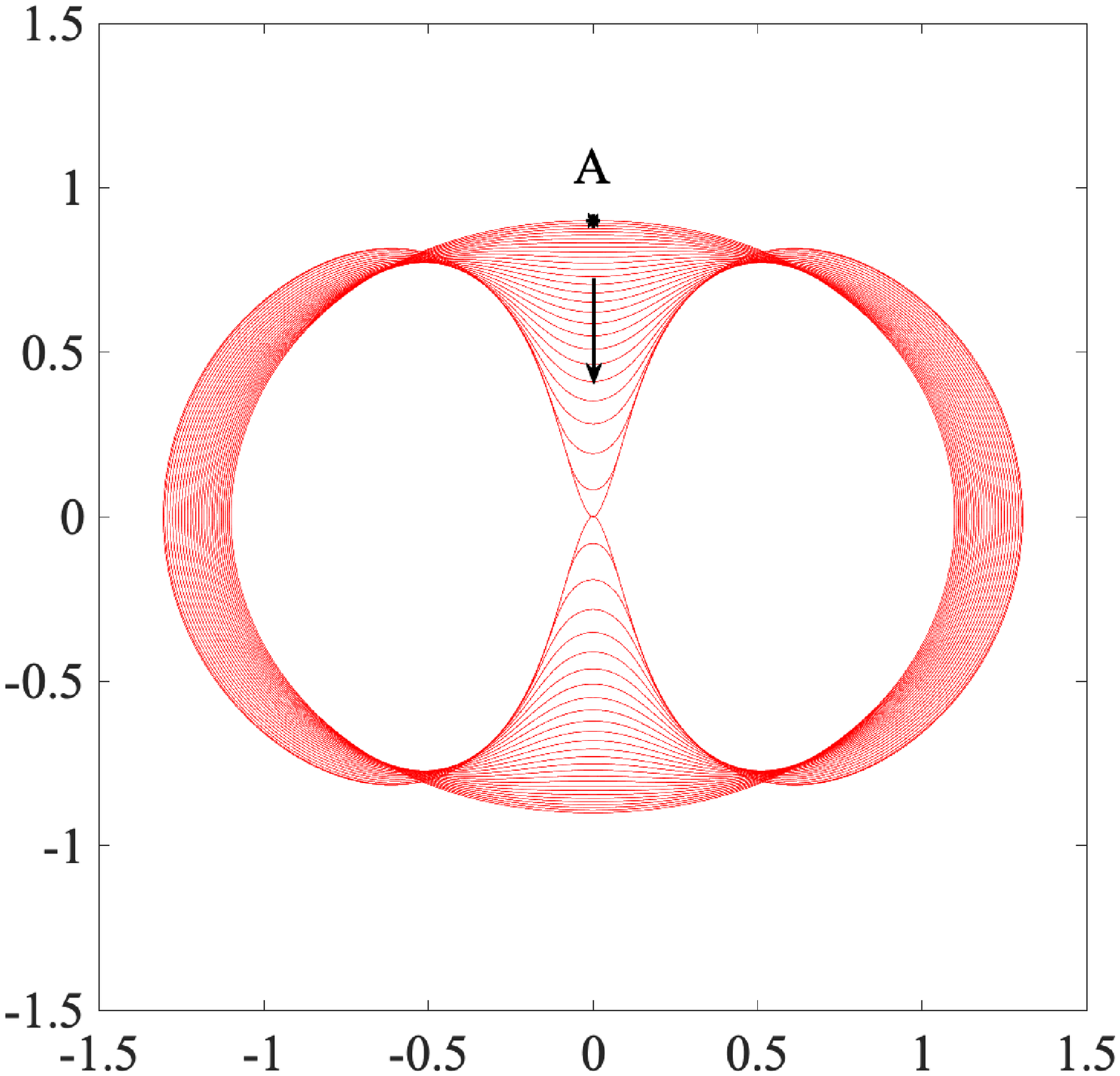}
\includegraphics[scale=0.3]{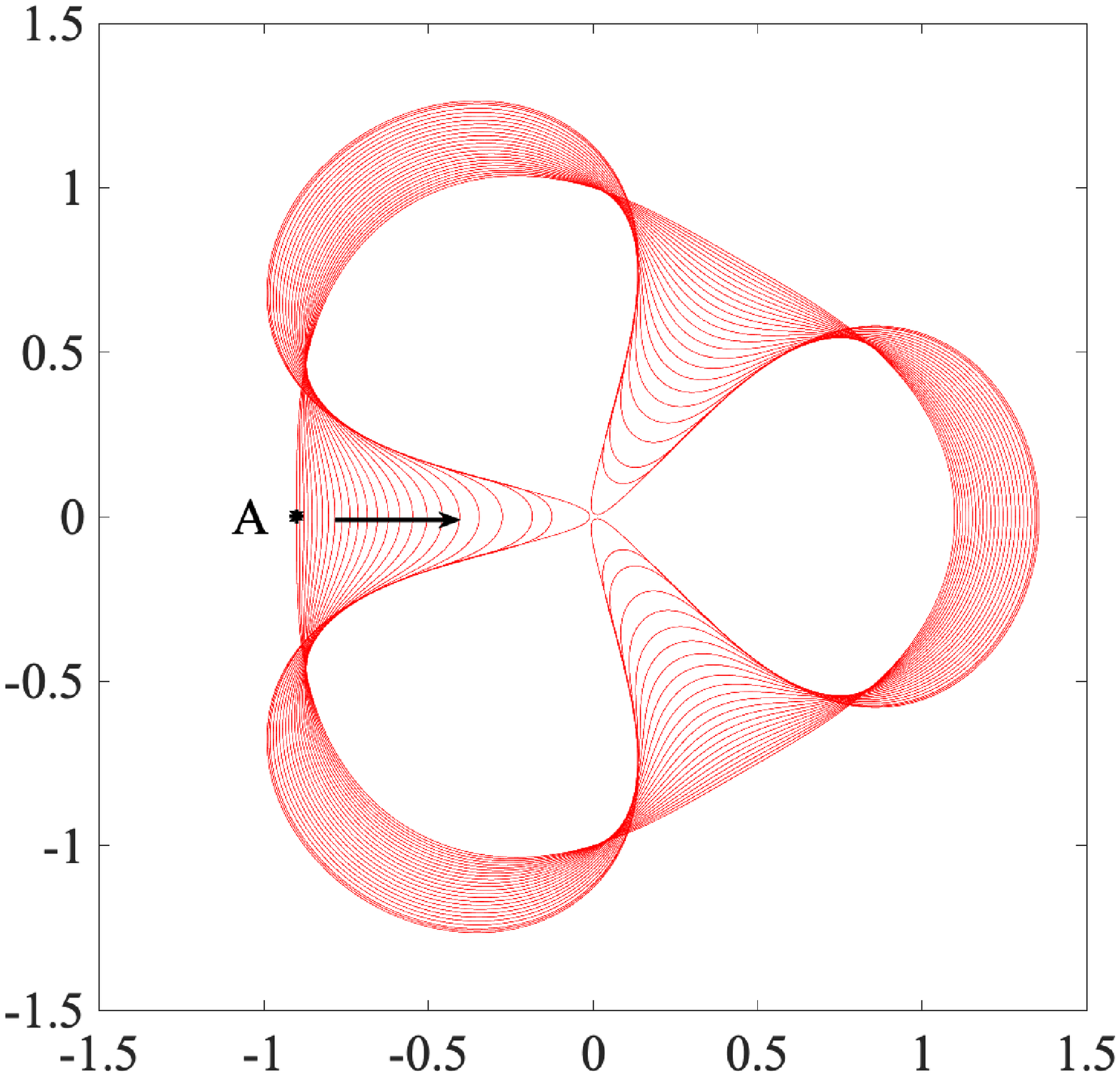}\\
\includegraphics[scale=0.3]{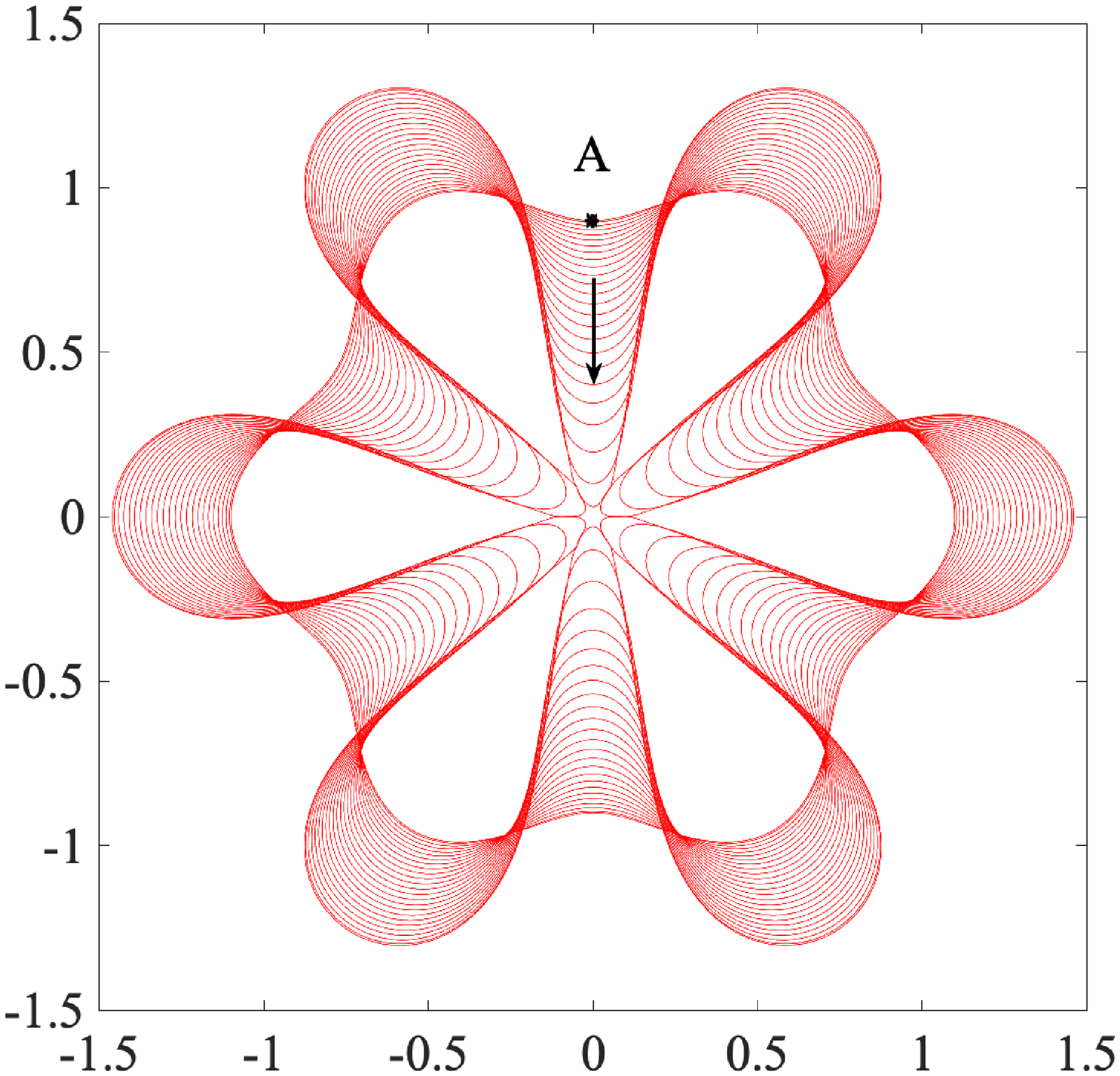}
\includegraphics[scale=0.3]{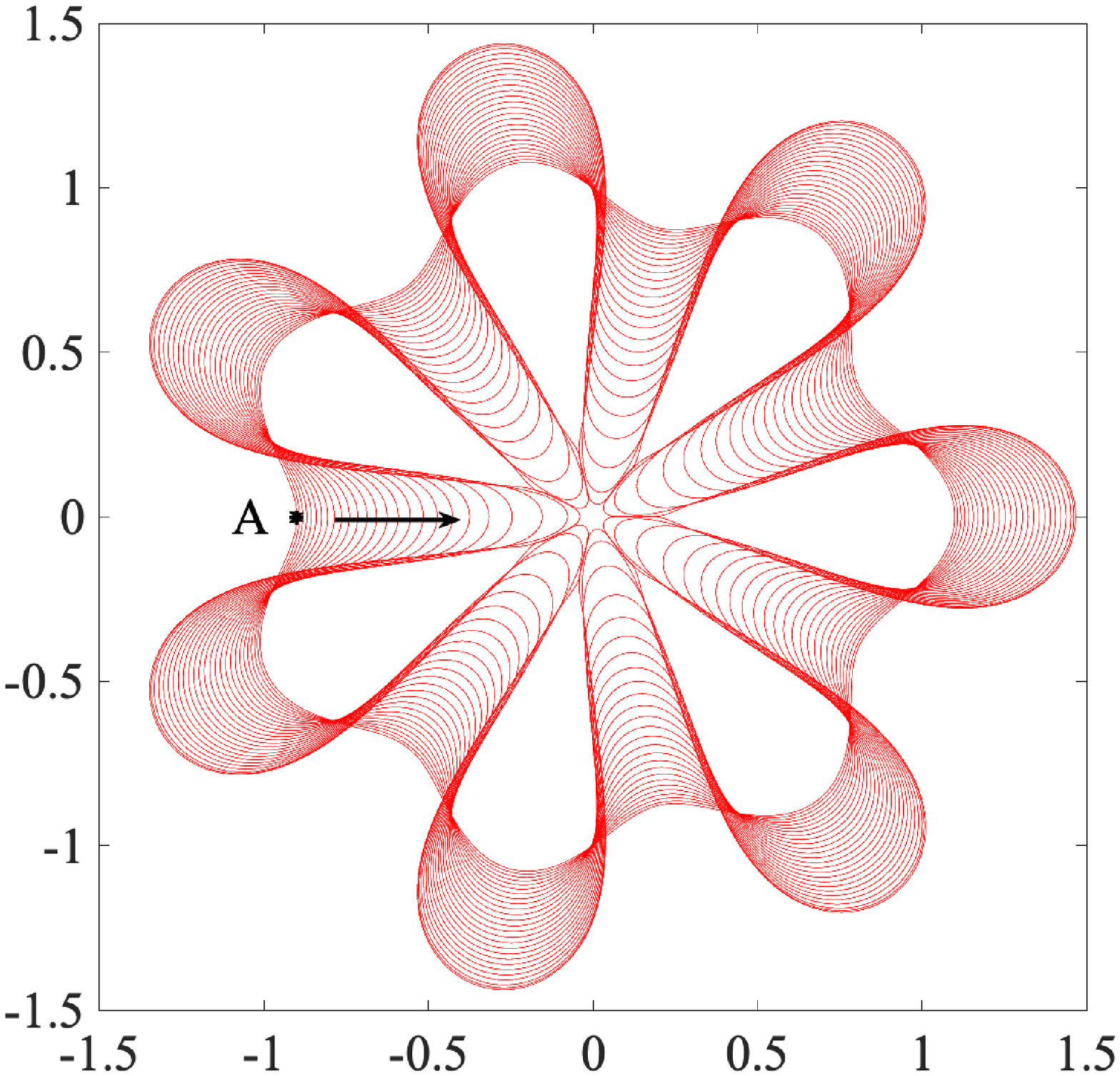}
\caption{Interface dynamics under zero fux $J=0$, constant current $I=-150I_0$ and surface tension $\tau=2.16\times 10^{-2}$. We take different  initial shapes $r(\theta,0)=1+0.1\cos(n\theta)$, where $n=2$, $3$, $6$, and $7$. The arrow in each plot indicates the evolution direction.}\label{Res:fig8}
\end{figure}

In this subsection, we explore the interface dynamics under zero flux, which means the area enclosed by the interface is conserved, and the system is unstable only due to electric effects. First we take a constant current $I=-150I_0$, surface tension $\tau=2.16\times 10^{-2}$, and various initial conditions $r(\theta,0)=1+0.1\cos(n\theta)$, where $n=2$, $3$, $6$, and $7$. The evolution of the interface are shown in \cref{Res:fig8}. The arrow indicates the evolution direction. We find that in all cases, the interface develops fingers whose bases tend to approach the origin. For lower modes $n=2$ and $3$, the interface does not develop thin tail region and tends to break into drops. On the other hand, the interface with higher modes $n=6$ and $7$ exhibits thin tail regions, connecting drops and an inner region. The inner region shrinks as the interface approaches the origin.

\begin{figure}[tbhp]
\includegraphics[scale=0.28]{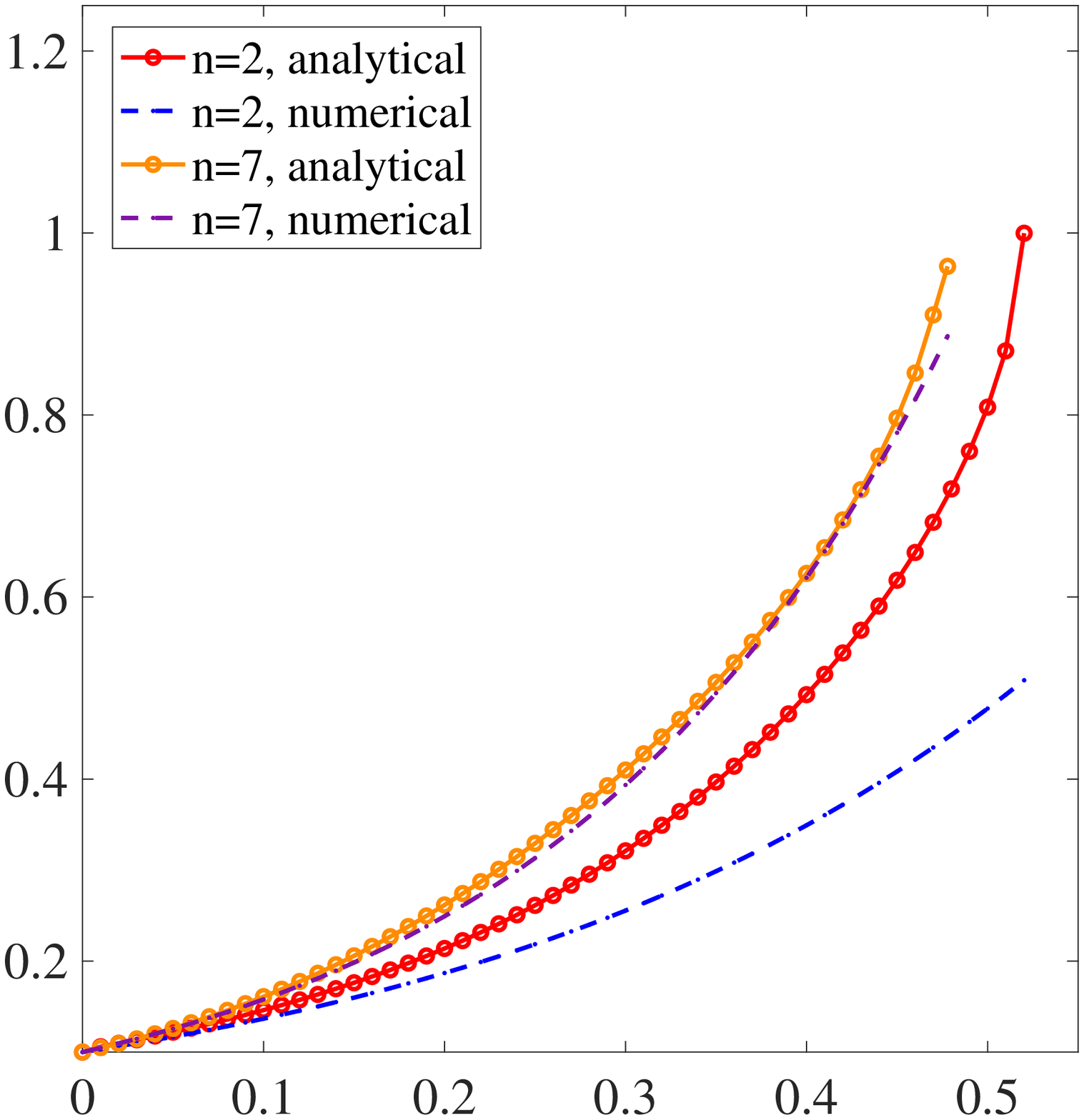}[a]
\includegraphics[scale=0.28]{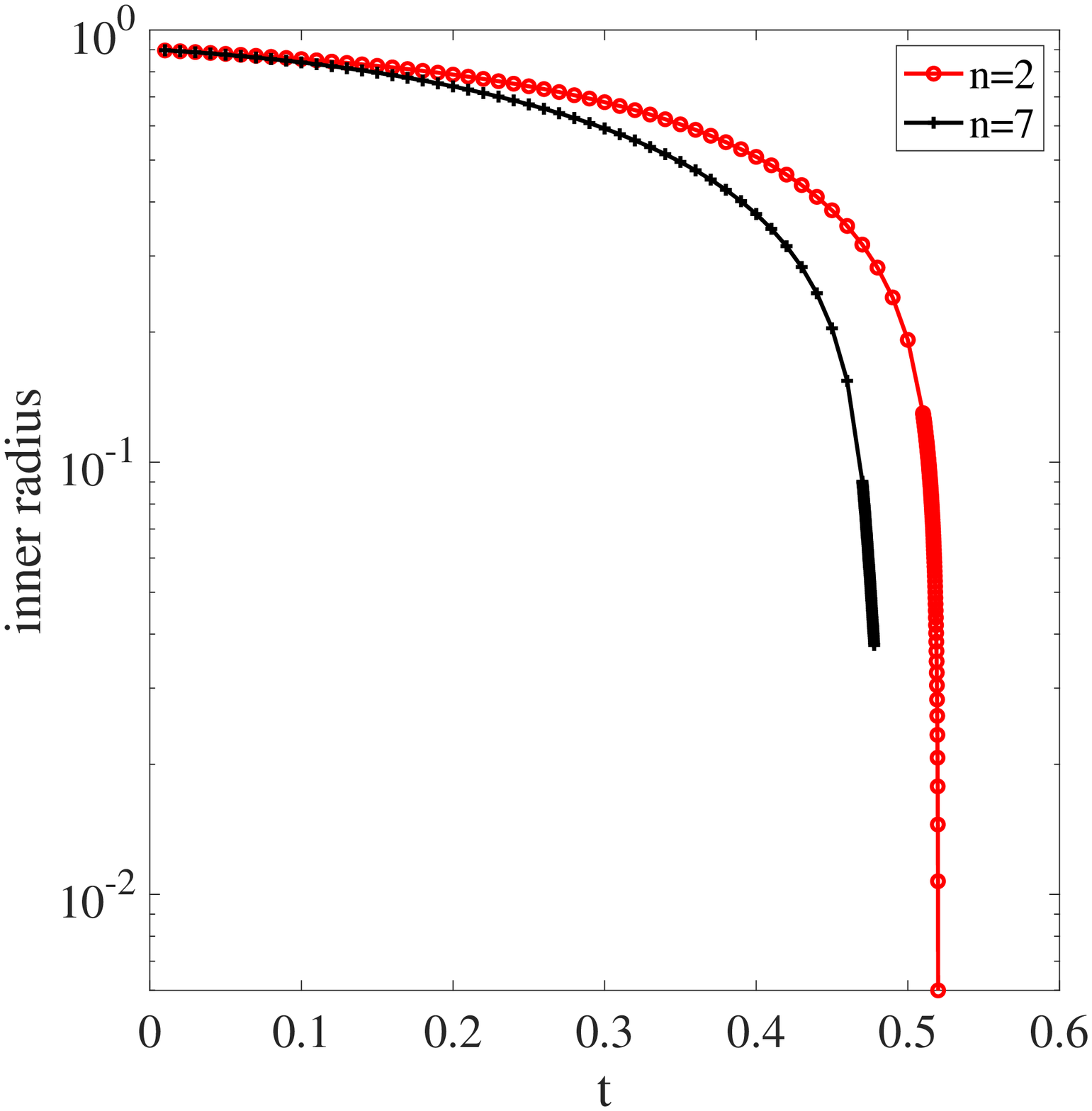}[b]
\includegraphics[scale=0.28]{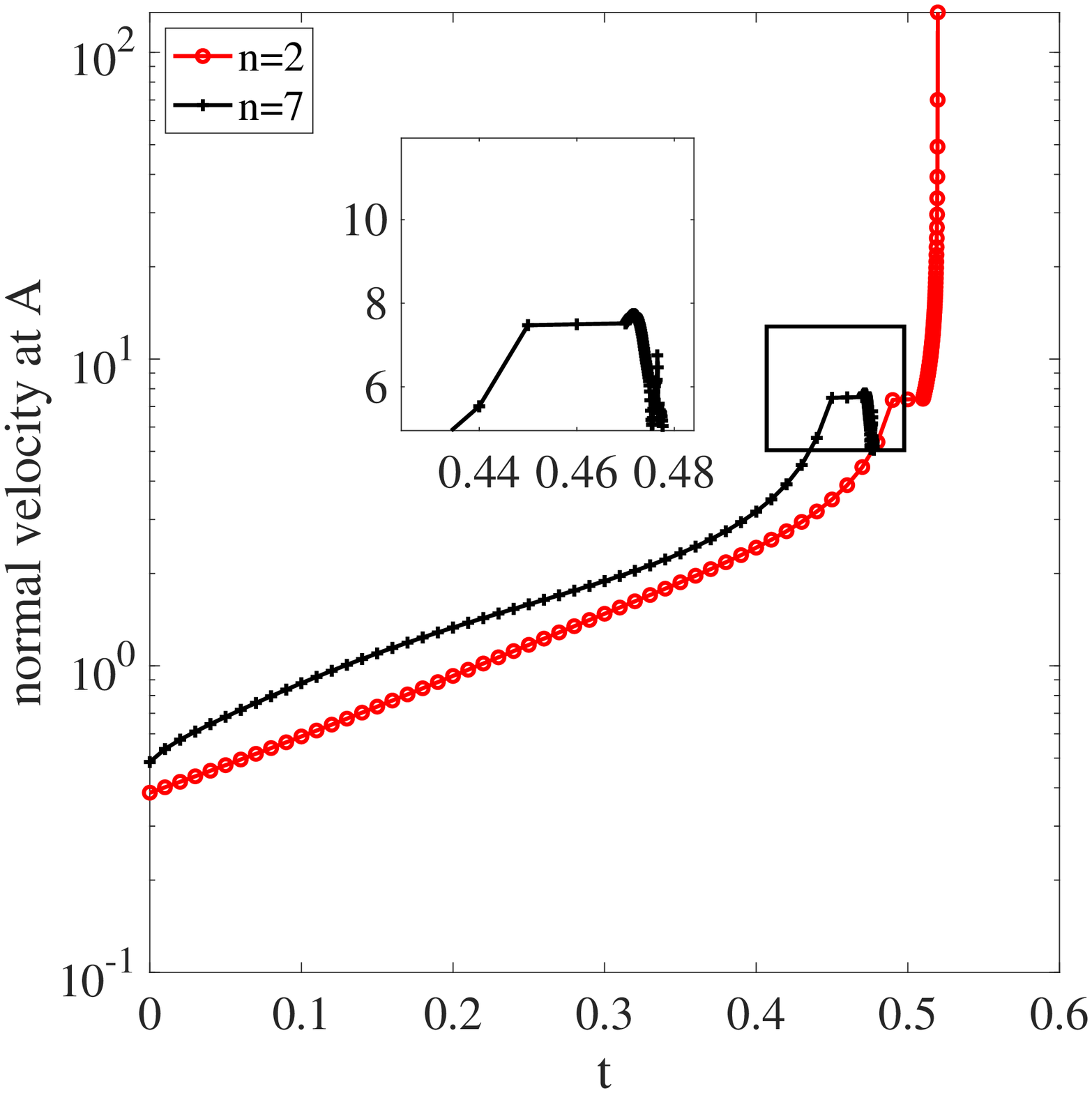}[c]
\caption{Properties of the interface in \cref{Res:fig8}. [a] shows the comparison between the linear theory and simulation results.  [b] shows the inscribed circle radius as a function of $R$. [c] shows the normal velocity of point A. Note that we take the opposite of the actual velocity in order to exhibit it in log scale.}\label{Res:fig9}
\end{figure}

\Cref{Res:fig9}[a] shows that linear theory expects an exponential behavior of the shape factor for $n=2$ and $7$, where the exponent depends on $n$. Our simulations reveal that for mode $n=2$, the shape factor is greater than linear prediction.
 For mode $n=7$, the shape factor from simulation is greater than linear results at very early times. Then it becomes less than linear expectation for a short period. And finally it grows above the theoretic value. In \cref{Res:fig9}[b], we present the inscribed circle radius for different $n$. Both radius tend to drop to zero at a finite time implying that the interface reaches the origin at a finite time. We find the radius obeys an algebraic law $\displaystyle (t_*-t)^b$, where $t_*$ is the estimated time when the interface reaches the origin. Analyzing the data, we find out that $t_*$  is about $0.520$ for $n=2$ and the exponent $b$ is $0.570$. For the higher mode $n=7$,  $t_*$ is $0.478$ and $b$  is $0.615$. Although both inner radius obey an algebraic law, the underlying physics are different. We  investigate the normal velocity at point A (indicated in \cref{Res:fig8}).  Note in \cref{Res:fig9}[c], we show the opposite value of the actual velocity, i.e, we multiply by negative one in order to exhibit it in log scale. A zoom-in of the box is shown as an inset. In both cases, the velocity increases in magnitude, indicating that point A moves inward faster as time elapses. For $n=2$, the velocity tends to blow up, while the velocity in $n=7$ case tends to oscillate around some finite values. These findings suggest that in all the cases, the interface tends to reach the origin, but it happens faster for the lower mode.
 
\begin{figure}[tbhp]
\includegraphics[scale=0.28]{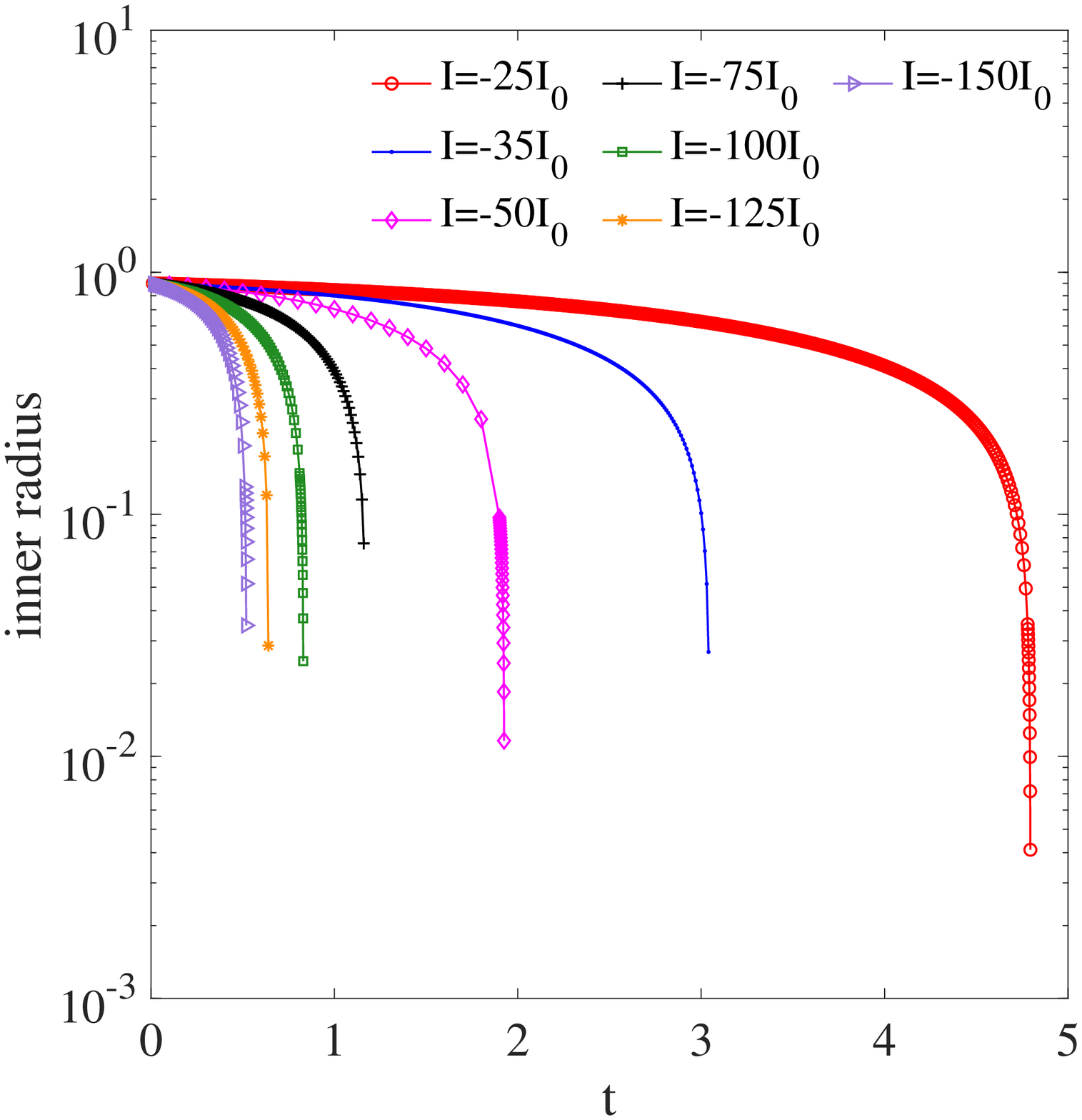}[a]
\includegraphics[scale=0.28]{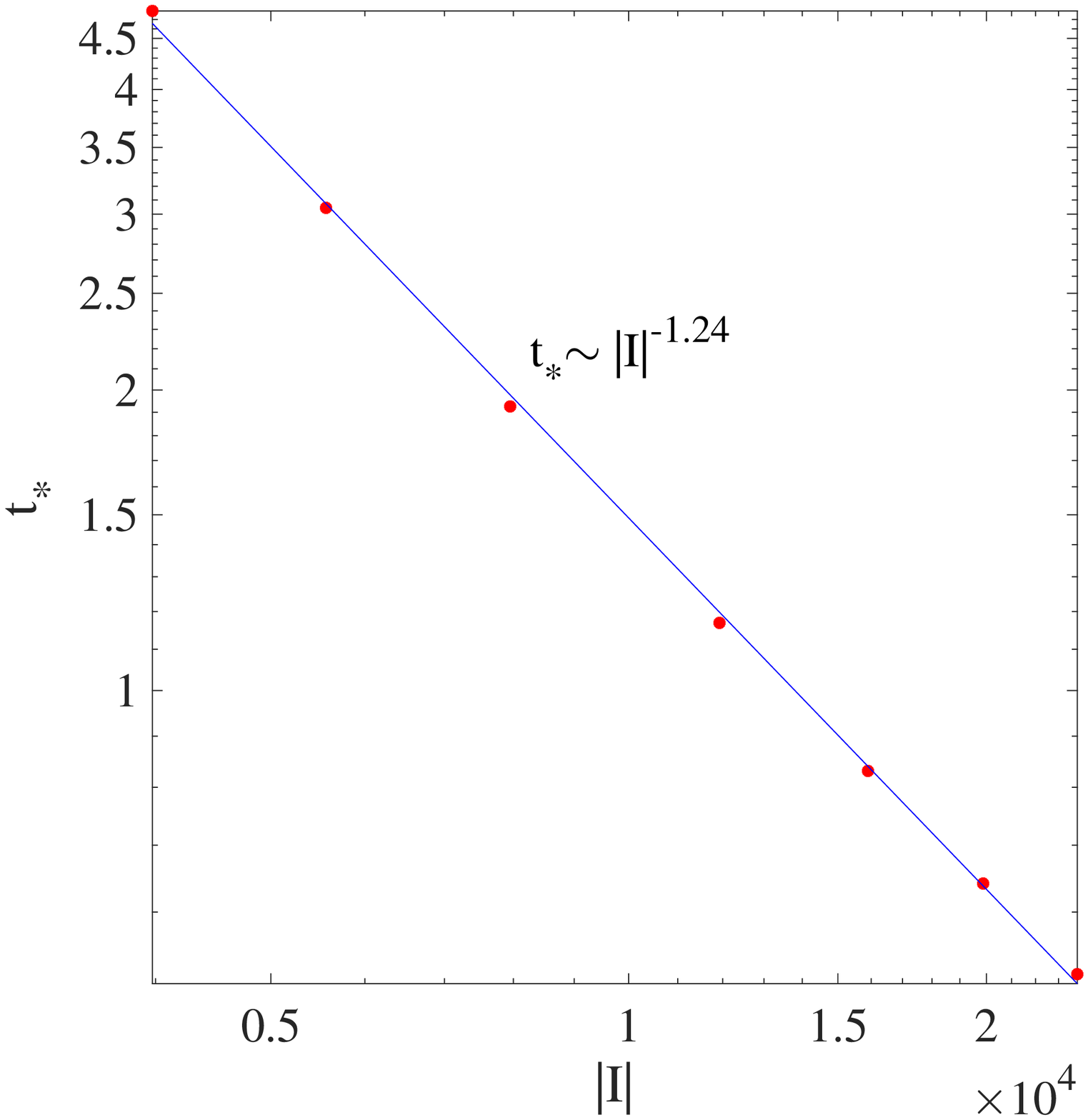}[b]
\caption{[a] shows the inscribed circle radius as a function of $t$ for different current $I$, where $J=0$,  surface tension $\tau=2.16\times 10^{-2}$, and  initial shape is $r(\theta,0)=1+0.1\cos(2\theta)$. They all obey the algebraic law $(t_*-t)^b$. [b] shows a loglog plot of estimated pinchoff time $t_*$ and absolute value of current $I$ in [a]. }\label{Res:fig10}
\end{figure}

Next we investigate the role of the current by fixing the surface tension $\tau=2.16\times 10^{-2}$ and initial shape $r(\theta,0)=1+0.1\cos(2\theta)$. We vary the current from $-25I_0$ to $-150I_0$. The interface tends to form two drops similar to those in \cref{Res:fig8}. In \cref{Res:fig10}[a], we present the inscribed circle radius as a function of time for different current. In all cases the radius obeys the algebraic law $\displaystyle (t_*-t)^b$, suggesting the interface reaches the origin at a finite time.  Fitting the data, we are able to estimate $t_*$ for each case. In \cref{Res:fig10}[b], we demonstrate the relationship between $t_*$ and current $I$, which reads as $t_* \sim |I|^{-1.24}$. 

\begin{figure}[tbhp]
\centering
\includegraphics[scale=0.38]{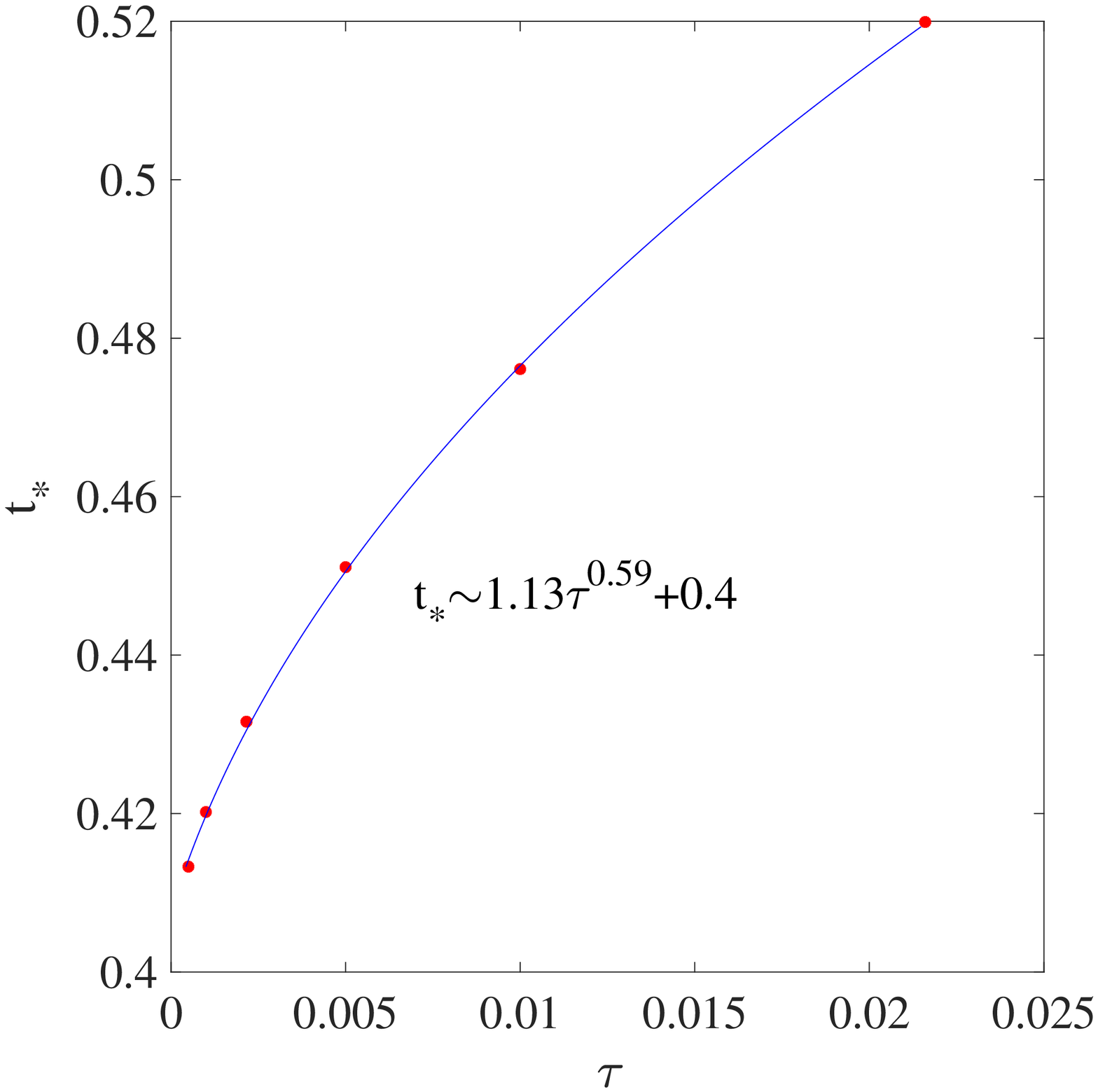}
\caption{It shows a loglog plot of estimated pinching time $t_*$ and surface tension $\tau$, when $J=0$, $I=-150I_0$ and initial shape $r(\theta,0)=1+0.1\cos(2\theta)$.}\label{Res:fig11}
\end{figure}

We also study the effect of surface tension via fixing the current $I=-150I_0$ and initial shape $r(\theta,0)=1+0.1\cos(2\theta)$.  We use various surface tensions from $2.16\times 10^{-2}$ to $5\times 10^{-4}$. The inscribed circle radius for different surface tension also obeys the algebraic law $\displaystyle (t_*-t)^b$.  We estimate $t_*$ for different surface tension and summarize the results in \cref{Res:fig11}.  It shows that $t_*$ decreases as $\tau$ decreases and satisfies $\displaystyle t_*\sim1.13\tau^{0.59}+0.4$. 

\subsection{self-similar shape}
According to linear theory, current $I_c$ satisfying \cref{GE:ssI}  makes the fastest growing $n_{max}$ fixed all the time. When no flux is inject, $I_c$ is a constant. As we discussed in previous subsections, the interface does not experience a self-similar shape but rather approaching the origin. When a constant flux is used, linear theory \cref{GE:ssg} predicts a stable interface. Our simulations confirm this phenomena (not shown here).  Thus, to produce a self-similar pattern, we focus on using the self-similar flux $J_d$ \cref{GE:ssJ}, which balances with the surface tension term in \cref{GE:ssg}, and the corresponding $I_c$. 

\begin{figure}[tbhp]
\centering
\includegraphics[scale=0.58]{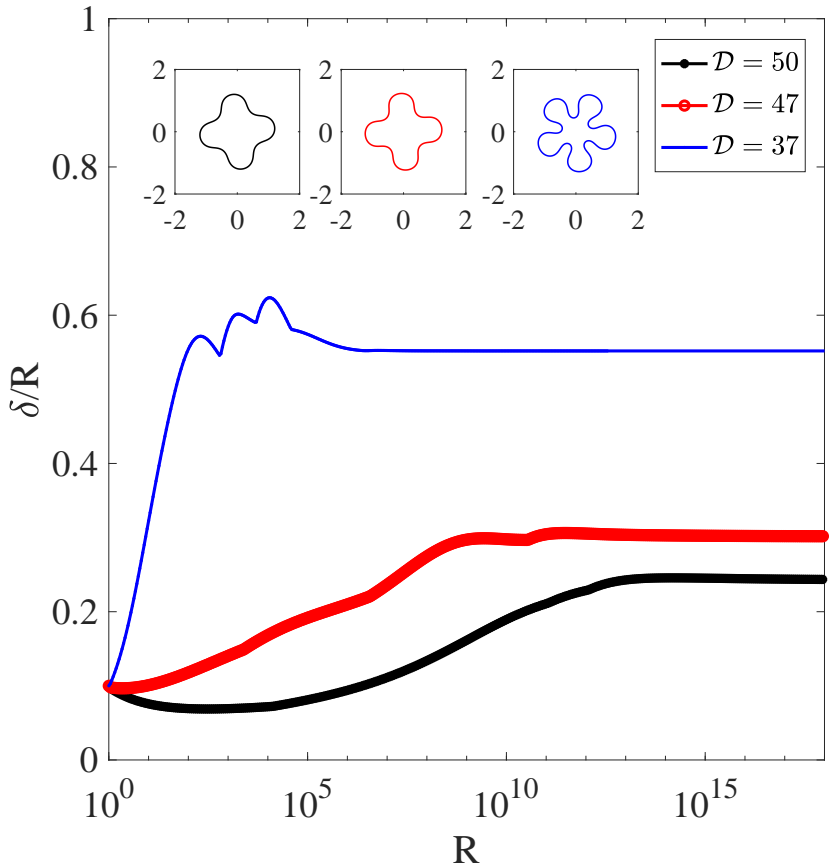}
\caption{It shows the shape factor under $I_c$ with $\mathcal C=47$, surface tension $\tau=2.16\times 10^{-2}$, and $J_d$ in \cref{GE:ssJ} with $\mathcal D=37$, $47$, and $50$. The initial shape is $r(\theta,0)=1+0.05(\sin(2\theta)+\cos(3\theta))$. The limiting self-similar shapes are shown as insets.}\label{Res:fig12}.
\end{figure}

In this subsection, we take the initial shape $r(\theta,0)=1+0.05(\sin(2\theta)+\cos(3\theta))$ and surface tension $\tau=2.16\times 10^{-2}$. We fixed $\mathcal{C}=47$ which means mode 4 has the fastest linear growth rate. The self-similar dynamics is studied for various flux. Specifically, we take $\mathcal{D}=50$, $47$, and $37$. In \cref{Res:fig12},  the nonlinear shape factor $\displaystyle (\frac{\delta}{R})_{NL}$ is plotted as a function of $R$. The final shapes of the interface are shown as insets. At early times, the shape factor grows dramatically, consistent with linear prediction. Later, nonlinear effects such as interactions and competitions between different modes stabilize the interface and lead to self-similar evolutions. Note that different choices of $\mathcal{D}$ lead to different self-similar patterns. The difference is not only in terms of the shape factor, but also in symmetry. For $\mathcal D=47$, the system is pure hydraulic and the interface develops a 4-fold self-similar pattern. For $\mathcal D=37$, $I_c$ is negative. That is the hydraulic part is not strong enough to maintain mode 4 to be the fastest growing mode. As a result, an electric term is needed to enforce growth. But our simulation demonstrates a 5-fold shape due to the nonlinear effects of this current and flux. For $\mathcal D=50$, $I_c$ is positive which means an electric term is used to suppress the growth.  Although the limiting shape is 4-fold, the interface shape is different. Our simulations reveal that parameters $\mathcal C$ and $\mathcal D$ together play an important role in selecting the limiting shape.

\section{Conclusions}
\label{sec:Con}

In this paper, we have investigated the dynamics of an expanding interface in a Hele-Shaw cell coupled with an electric field. When a less viscous fluid invades a more viscous fluid, the interface develops the well-known Saffman-Taylor instabilities. According to linear theory, the instability is the balance between the driving term from the flux $J$, destabilizing term from surface tension $\tau$, and the current $I$. Note that current is able to make the interface stable or unstable depending on the direction of current. We have demonstrated that mode $n$ possesses the fastest linear growth rate when a special current $I_c$ in \cref{GE:ssI} is applied. In the linear analysis, there is only one control strategy either in terms of current $I$ or flux $J$. 

To simulate the nonlinear dynamics of the interface, we have developed an efficient, highly accurate boundary integral method which utilizes space and time rescaling to track the interface. We have accelerated the slow dynamics of the interface by rescaling the origin time and space to a new frame which maintains the area enclosed by the interface. This enables us to study the long-time dynamics of the interface. Our results reveal that the method is efficient and spectrally accurate in space, and they are in good agreement with the analytical solution. 

We have exploited the evolution of interface morphologies to very long times under a constant flux $J=1$ and different currents. Comparing the morphologies, we have found that a positive current prefers to suppress the repeated tip-splitting process and promote fingers with similar size. On the other hand, a negative current stimulates the interfacial instability and leads to a thin tail on the interface which connects the finger and a small drop of the inner fluid at the center. Using a small surface tension, the tail region disappears since the unstable effects are too strong. 

Another interesting point is the possibility of the interface topological changes. We have investigated the interface morphologies under a zero flux and different negative currents. In general, the interface tends to reach the origin and the inner fluid may break into several drops. We have found the smallest distant from interface to the origin obeys an algebraic law $\displaystyle (t_*-t)^b$, where $t_*$ is the pinchoff time and depends on the current $I$, surface tension $\tau$, and the initial mode $n$.  To clearly understand this phenomenon, an analytical study may be necessary for future work.

At last, we have examined the nonlinear self-similar shape of the interface. Using the special current $I_c$ in \cref{GE:ssI} and the self-similar flux $J_d$ in \cref{GE:ssJ}, our simulations show that there exist self-similar patterns, which is the combination of the interactions and competition between different modes. In other words, the interfacial instability is saturated by nonlinear effects. We also have found that the current and the flux both control the self-similar pattern.

%\appendix
%\section{An example appendix} 
%\lipsum[71]
%
%\begin{lemma}
%Test Lemma.
%\end{lemma}

\section*{Acknowledgments}
W. Y. thanks the support from the National Science Foundation of China grants DMS-11771290. S. L. and J. L. acknowledge the support from the National Science Foundation, Division of Mathematical Sciences (NSF-DMS) grants DMS-1714973, 1719960, 1763272 (J. L.) and DMS- 1720420 (S. L.). J. L. thanks the support from the Simons Foundation (594598QN) for a NSF-Simons Center for Multiscale Cell Fate Research. J. L. also thanks the National Institutes of Health for partial support through grants 1U54CA217378-01A1 for a National Center in Cancer Systems Biology at UC Irvine and P30CA062203 for the Chao Family Comprehensive Cancer Center at UC Irvine.
\bibliographystyle{siamplain}
\bibliography{references}
\end{document}